%% file: main.tex
\documentclass{sig-alternate} 
\usepackage{mathptmx} 

\newcommand{\ignore}[1]{}
\usepackage{fancyhdr}
\usepackage[normalem]{ulem}
\usepackage[hyphens]{url}
\usepackage{microtype}
\usepackage{ascmac}
\let\Return\undefined 
\usepackage{fancybox}
\usepackage{algorithm}
\usepackage{algpseudocode}
\usepackage{caption}
\usepackage{subcaption}
\usepackage{wrapfig}
\usepackage[braket]{qcircuit}
\newcommand{\maj}{\txt{M\\A\\J}}
\newcommand{\mmaj}{\multigate{2}{\maj}}
\newcommand{\gmaj}{\ghost{\maj}}
\newcommand{\uma}{\txt{U\\M\\A}}
\newcommand{\muma}{\multigate{2}{\uma}}
\newcommand{\guma}{\ghost{\uma}}

\algrenewcommand\algorithmicindent{1.0em}%

\usepackage[bookmarks=true,breaklinks=true,letterpaper=true,colorlinks,linkcolor=black,citecolor=blue,urlcolor=black]{hyperref}

\fancypagestyle{firstpage}{
  \fancyhf{}

  \fancyhead[C]{}
  \fancyfoot[C]{\thepage}
}  

\title{Extracting Success from IBM's 20-Qubit Machines Using Error-Aware Compilation}
\numberofauthors{5}
\author{
\alignauthor
Shin Nishio\titlenote{These authors contributed equally to this work.}\\
  \affaddr{Faculty of Policy Management}\\
  \affaddr{Keio University}\\
  \affaddr{5322 Endo, Fujisawa City}\\
  \affaddr{Kanagawa, Japan}\\
  \email{parton@sfc.wide.ad.jp}
\alignauthor
Yulu Pan\raisebox{9pt}{$\ast$}\\
  \affaddr{Faculty of Science and Technology}\\
  \affaddr{Keio University}\\
  \affaddr{3-14-1 Hiyoshi, Kohoku-ku}\\
  \affaddr{Yokohama, Japan}\\
  \email{pandaman@am.ics.keio.ac.jp}
\alignauthor
Takahiko Satoh\\
  \affaddr{Quantum Computing Center}\\
  \affaddr{Keio University}\\
  \affaddr{3-14-1 Hiyoshi, Kohoku-ku}\\
  \affaddr{Yokohama, Japan}\\
  \email{satoh@sfc.wide.ad.jp}
\and
\alignauthor
Hideharu Amano\\
  \affaddr{Faculty of Science and Technology}\\
  \affaddr{Keio University}\\
  \affaddr{3-14-1 Hiyoshi, Kohoku-ku}\\
  \affaddr{Yokohama, Japan}\\
  \email{hunga@am.ics.keio.ac.jp}
\alignauthor
Rodney Van Meter\\
  \affaddr{Faculty of Environment and Information Studies}\\
  \affaddr{Keio University}\\
  \affaddr{5322 Endo, Fujisawa City}\\
  \affaddr{Kanagawa, Japan}\\
  \email{rdv@sfc.wide.ad.jp}
}

\begin{document}
\maketitle 
\thispagestyle{firstpage}
\pagestyle{plain}


\begin{abstract}
  NISQ (Noisy, Intermediate-Scale Quantum) computing requires error
  mitigation to achieve meaningful computation.  Our compilation tool
  development focuses on the fact that the error rates of individual
  qubits are not equal, with a goal of maximizing the success
  probability of real-world subroutines such as an adder circuit.  We
  begin by establishing a metric for choosing among possible paths and
  circuit alternatives for executing gates between variables placed
  far apart within the processor, and test our approach on two IBM
  20-qubit systems named Tokyo and  Poughkeepsie.  We find that a single-number metric
  describing the fidelity of individual gates is a useful but
  imperfect guide.

  Our compiler uses this subsystem and maps complete circuits onto the
  machine using a beam search-based heuristic that will scale as
  processor and program sizes grow.  To evaluate the whole compilation
  process, we compiled and executed adder circuits, then calculated
  the KL-divergence (a measure of the distance between two probability
  distributions).  For a circuit within the capabilities of the
  hardware, our compilation increases estimated success probability
  and reduces KL-divergence relative to an error-oblivious placement.
\end{abstract}

\input{intro}
\input{related}

\input{cost}
\input{algorithm}
\input{conclusion}
\input{acknowledgement}


\clearpage
\newpage
\bibliographystyle{ieeetr}
\bibliography{ref,pandaman}

\end{document}

%% file: intro.tex
\section{Introduction}

Quantum computers
exist~\cite{grumbling18:quantum-natacad,ladd10:_quantum_computers,kandala2017hardware,zhang2017observation,bernien2017probing,barends2014superconducting,martin2012experimental},
and once mature, they will surpass classical computers on a range of
important
problems~\cite{preskill:nisq,Ronnow25072014,bennett:strengths,aaronson:thesis,harrow2017quantum,mosca2008quantum,bacon10:_recent_progress,montanaro2015:qualgo-qi,2017npjQI...3...15L}.
Experimental progress in recent years has been rapid, with systems of
up to 20 qubits now accessible, and systems ranging from 49 qubits to
128 qubits either undergoing testing in the laboratory or promised for
the near future.  Quantum machines will overtake classical ones
somewhere between 50 and 150 qubits as quantum capability (especially
fidelity of gate operations) and improving classical simulation
techniques~\cite{markov:contracting-siam,pednault:1710.05867} compete,
first for demonstration problems then inevitably (we believe) for
problems of practical import.  This leaves computer engineers with
challenges in
architecture~\cite{oskin:quantum-wires,thaker06:_cqla,isailovic06:_interconnect,van-meter05:_distr_arith_quant_multic,van-meter13:_blueprint,van-meter16:_ieee-comp}
and programming
tools~\cite{gay05:_quant_progr_lang,green2013quipper,abhari12:_scaffold,JavadiAbhari:2014:SFC:2597917.2597939,wecker2014liqui,heckey2015compiler}.
In particular, because full realization of quantum error
correction~\cite{devitt13:rpp-qec,gottesman2009introduction,RevModPhys.87.307}
remains out of reach, in the near term, we must create error-aware
compilers for the noisy, intermediate-scale quantum computing
era~\cite{preskill:nisq}.

\begin{figure}[htbp]
    \begin{center}
        \includegraphics[width=8cm]{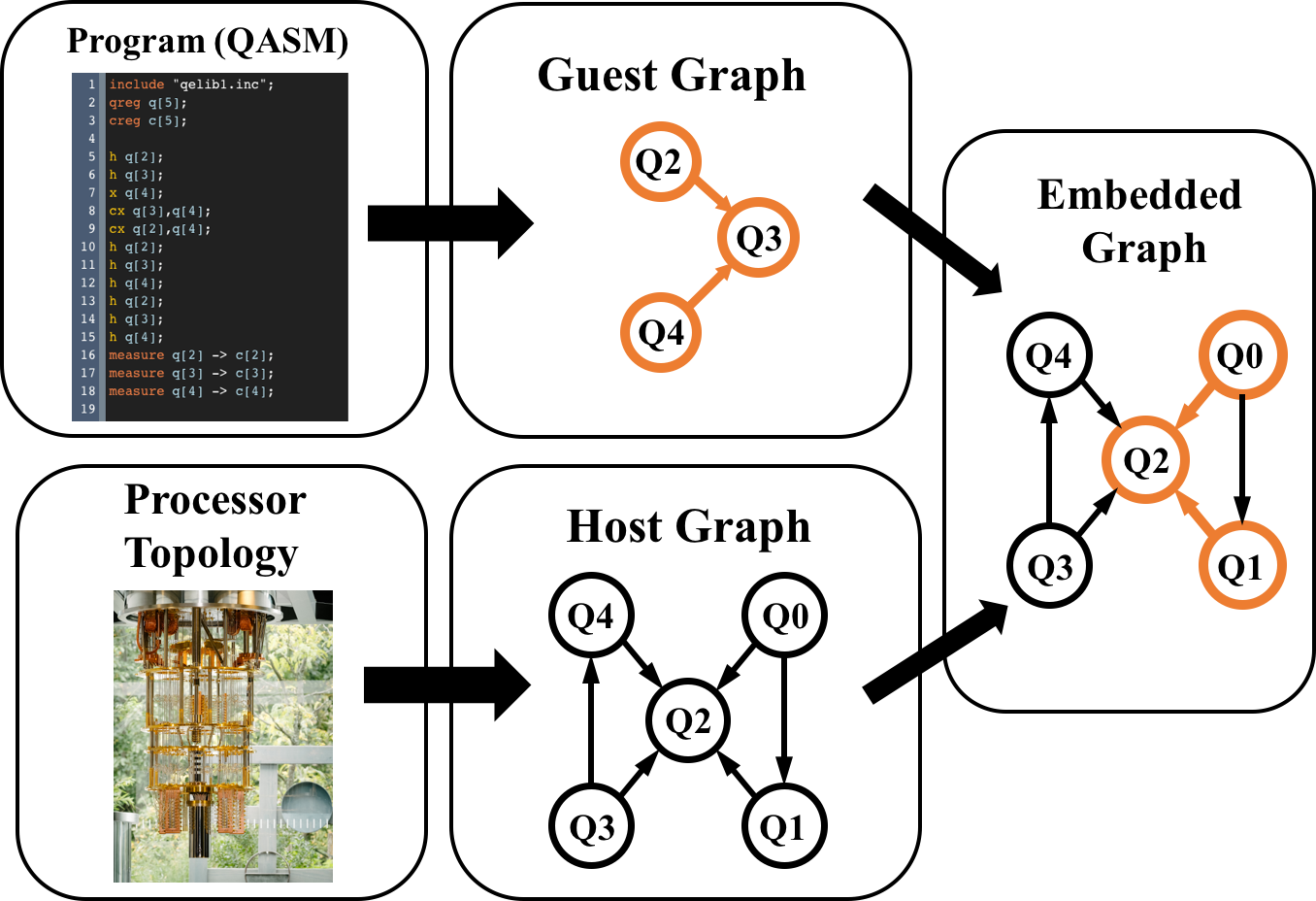}
        \caption{Circuit mapping: Gates between qubits in a program
          can be represented as edges in a graph.  The constraints of
          the physical system also can be represented in a graph.
          Compilation embeds the program (guest graph) in the physical
          topology (host graph) as it assigns qubit variables to
          locations within the machine.}
        \label{fig:mapping}
    \end{center}
\end{figure}

Compilation varies significantly depending on whether we are compiling
for fault-tolerant execution on top of error-corrected logical qubits,
or for the ``bare metal'' machine, and further whether qubits fly
(photons), can be moved modest distances (ions), or stay in place
(solid-state), and what connectivity constraints are
incurred~\cite{van-meter13:_blueprint,van-meter16:_ieee-comp,PhysRevX.2.031007}.
QEC compilation has been the subject of dramatic advances in recent
years
(e.g.,~\cite{selinger2015efficient,PhysRevA.87.022328,fowler2018low}),
but in this paper we focus on bare metal machines, which imposes a
different set of goals and constraints.  Machine-level compilation
involves a series of phases: first, decomposition of higher-level
language constructs into a series of one- and two-qubit operations
that can be executed on the target system; second, mapping of the
variables defined by the programmer to locations in the system, in
tandem with generation of appropriate execution of gates between
qubits unfortunately placed far apart; third, generation of low-level
control for the hardware itself.  This paper focuses on the second
phase, as shown in Fig.~\ref{fig:mapping}.  We test our ideas on a
specific system, the IBM 20-qubit machine named Tokyo~\footnote{Tokyo
  is only the name of the machine, it actually resides in Yorktown
  Heights.} and Poughkeepsie, but expect that the ideas will hold for a broad range of
solid-state systems.

\begin{figure}[htbp]
    \begin{center}
        \includegraphics[width=6cm]{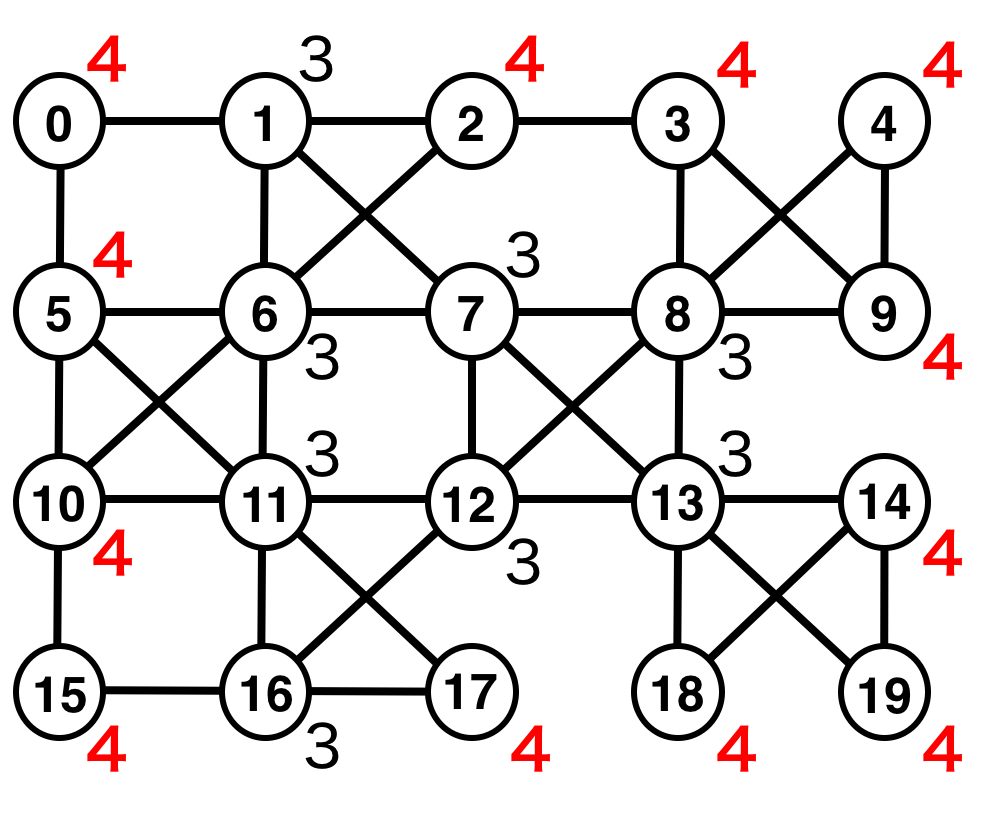}
        \caption{Architecture of IBM's 20-qubit processor named
          Tokyo. Each vertex represents a qubit, and each edge
          indicates that a CNOT gate can be executed between the two
          qubits.  The numbers outside the circles indicate the
          eccentricity, or maximum distance to another qubit.}
        \label{fig:IBMQ20}
    \end{center}
\end{figure}

\begin{figure}[htbp]
    \begin{center}
        \includegraphics[width=6cm]{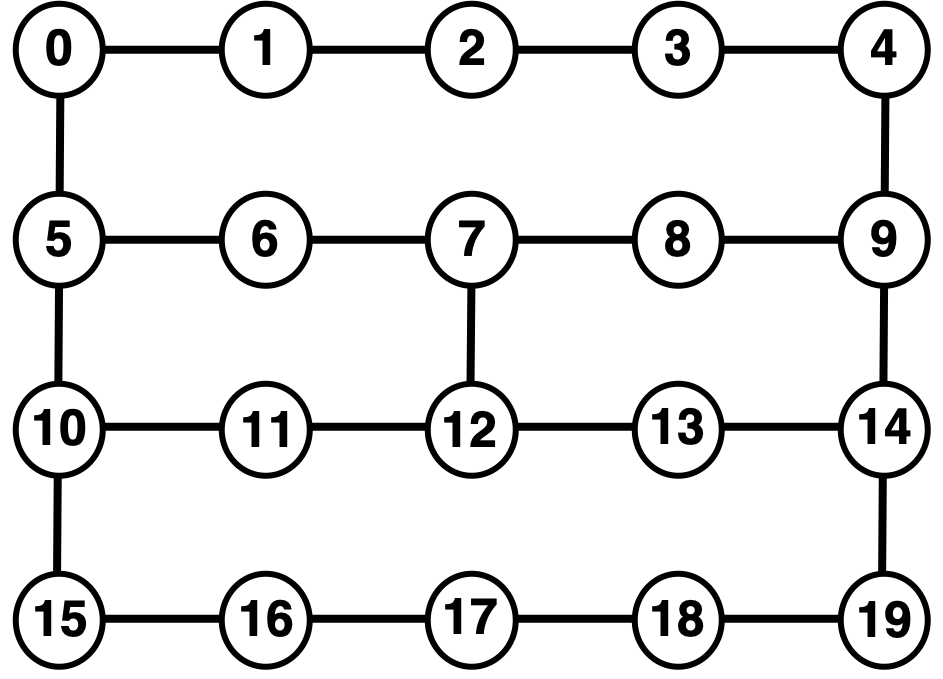}
        \caption{Architecture of IBM's 20-qubit processor named Poughkeepsie. Comparing Tokyo and Poughkeepsie, the number of connections between qubits that can execute CNOT is greatly reduced, but the gate fidelity is improved. as shown in  Table \ref{tab:specification}}
        \label{fig:IBMQ20_po}
    \end{center}
\end{figure}

Tokyo and Poughkeepsie are  transmon-style superconducting systems with limited
connectivity between qubits, as shown in
Fig.~\ref{fig:IBMQ20}~\cite{koch:PhysRevA.76.042319,ibm:qexp} and Fig.~\ref{fig:IBMQ20_po}.  Each
vertex indicates a qubit, and each edge indicates whether or not a
multi-quantum bit gate is physically executable.  Existing solid-state
quantum processors such as this have limitations on the execution of
multi-qubit gates such as CNOT because effects such as crosstalk make
it undesirable to cross qubit-to-qubit couplers or control wires or to
make all-to-all shared buses.

In existing quantum processors, fidelity degrades due to various
errors, affecting the success probability of calculations.  The
susceptibility to errors of each quantum bit on the quantum processor
can vary dramatically.  For Tokyo, two-qubit errors dominate.  The
error rates reported by the Qiskit tools do not distinguish between
bit flip errors and phase flip errors (the two primary types of
quantum state errors), giving us only a single number to work with.
The edges in Fig.~\ref{fig:IBMQ20} have reported error rates ranging
from 3\% to 12\%.  (Newer designs for qubit couplers may push those
errors to 1\% or below~\cite{roth:PhysRevA.96.062323}.)  Executing
medium-length programs on such a system is challenging, making it
imperative not only to minimize the total number of gates, but also to
assign program variables to qubits with a careful eye toward which edges will
be most used.  While a number of projects have focused on gate
reduction, we choose to make execution success probability our primary
goal.

To understand the system, we began by conducting a form of system
testing known as randomized benchmarking (RB) (described with other
background material in Sec.~\ref{sec:background}).  We omit our RB
results here because we chose to develop our compilation algorithms
using the error rates reported by the Qiskit tools (which are also
RB-derived), since extensive testing of the machine itself before
every application compilation is inherently impractical.

Our first task is to \emph{assess our ability to accurately predict
  the success rate} of a given circuit using the product of the
individual gate success probabilities as our \emph{estimated success
  probability} (ESP) (Sec.~\ref{sec:score}).  We find that our exact
numbers are off, but we can correctly choose which of two circuits
will be \emph{better} on the real machine about two-thirds of the
time.  We use this to choose how to execute long-distance gates across
the chip, selecting from among a group of circuits that are all
theoretically equivalent but in practice demonstrate large
differences.

The second task is to build on this capability and compile and test
complete circuits (Sec.~\ref{sec:qopter}).  Unfortunately, we know
that the circuit mapping phase, similar to place-and-route in hardware
design, is NP-complete~\cite{Siraichi:2018:QA:3179541.3168822}.
Therefore, we use a beam search-based heuristic that includes some
stochastic behavior.  We compile an adder
circuit~\cite{quant-ph/0410184} for input register sizes of one, two,
and four qubits, consisting of tens to well over a hundred two-qubit
gates.  Compilations are repeated with different random number seeds
to assess the performance.  For the smaller circuits, our compiler
results in executions on the quantum hardware with substantially less
divergence from the expected output distributions than simple random
placement of variables on the chip.  The largest circuits exceed the
current capabilities of the system.

We conclude from this data from a real quantum computer that the
simple ESP can be used to improve the success probability of quantum
computations for a reasonable (classical) compilation cost, and we
expect that this will scale to the largest processors projected to
exist in the next few years.  We suggest that further work on more
nuanced metrics will allow still further improvements
(Sec.~\ref{sec:conclusion}).

%% file: related.tex
\section{Background}
\label{sec:background}

Quantum computing research has been appearing in computer architecture
venues for over fifteen years, so we dispense with a complete
introduction
here~\cite{oskin:quantum-wires,thaker06:_cqla,isailovic06:_interconnect,van-meter05:_distr_arith_quant_multic}.
Instead, we wish to focus on the key problems with measuring and
modeling errors to achieve high success rates in execution, and their
relationship to compilation that will lead to using quantum computers
to solve problems that classical systems cannot.

\subsection{Error model}

\begin{figure}[htbp]
    \centering
    \includegraphics[width=6cm]{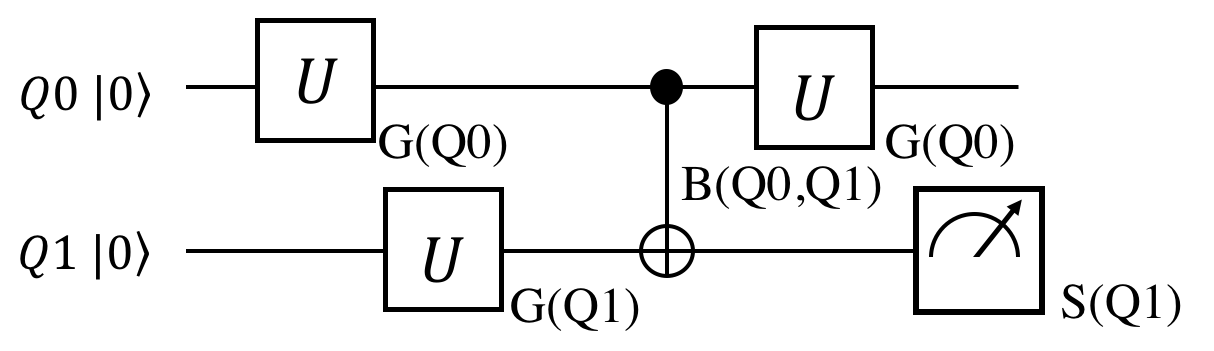}
    \caption{Errors and the gates with which they occur}
    \label{error}
\end{figure}

Our model classifies errors in quantum circuits into three groups.
Fig.~\ref{error} shows where such errors occur.  Rather than a formal
quantum mechanical model~\cite{lidar13:_intro_to_decoh_and_noise},
here we need only the error probabilities.

\begin{enumerate}
\item Single-Qubit Gate errors ($G$) may be unitary bit flips
  ($\ket{0}\rightarrow\ket{1}$ and $\ket{1}\rightarrow\ket{0}$) or
  phase flips
  ($(\alpha\ket{0}+\beta\ket{1})\rightarrow(\alpha\ket{0}-\beta\ket{1})$),
  where $\{\alpha,\beta\}\in\mathbb{C}$ are the complex amplitudes of
  a qubit's state, or they may be non-unitary errors such as
  relaxation, in which \ket{1} tends to decay to the lower-energy
  state \ket{0};
\item Bi-Qubit gate errors (CNOT error) ($B$) may also flip the value
  or phase of one or both qubits, and are particularly insidious
  because they \emph{propagate} errors from one qubit to another; they
  are critical to creating quantum entanglement and appear in
  algorithms in large numbers; and
\item State Preparation And Measurement (SPAM) errors ($S$) are
  important but can appear only once per qubit in a program execution
  on the IBM machines, and so have less cumulative impact on success.
\end{enumerate}

IBM publishes the value of each of these errors on a per-qubit or
per-coupler basis as a backend property of each machine, retrievable
using a Qiskit call available to IBM Q Network members.  Tokyo is
calibrated once a day, using a procedure called \emph{randomized
  benchmarking}, below.

Beyond these straightforward errors, various crosstalk and resonances
within the system cause the state of qubits to affect one another.  A
complete characterization of this would require extensive
\emph{tomography}.

\subsection{Tomography}

Quantum tomography comes in two primary forms: state tomography and
process tomography.  This tomography tells us how well we have done at
creating our desired state, or how well a particular process (gate or
set of gates) works, respectively.  It can be used to characterize
errors.  However, a practical problem arises: the number of possible
states naturally grows exponentially with the number of qubits;
moreover, we need to test not just for bit flip errors, but also for
phase flip errors.  Ultimately, we may need to execute the creation of
the state or the gate sequence $k3^n$ times for $n$ qubits, where $k$
is a constant determined by the precision we require for the
reconstruction and may be thousands.  Although only tomography or a
similarly rigorous (and heavyweight) procedure can tell us about the
state-dependent crosstalk and other factors, it is impractical even at 20 qubits and altogether beyond reach for larger systems.  This need led to the creation of randomized benchmarking.

\subsection{Randomized Benchmarking}

\begin{figure}[htbp]
    \centering
    \includegraphics[width=6cm]{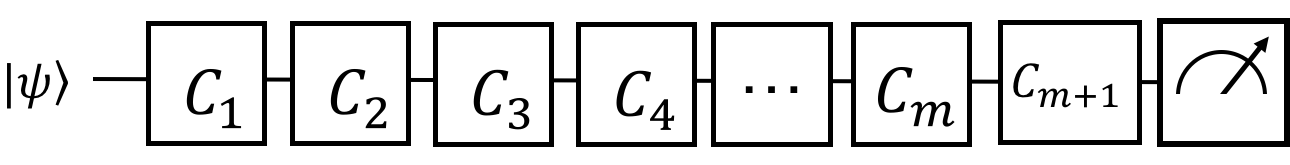}
    \caption{Randomized Benchmarking}
    \label{fig:rb}
\end{figure}

We can assess the fidelity of a set of possible gates (e.g., a
commonly used set of gates known as the \emph{Clifford group}) under a
broad range of conditions with dense coverage of input states by
using the following randomized benchmarking (RB)
procedure~\cite{knill:PhysRevA.77.012307,magesan:PhysRevA.85.042311}.
Fig.~\ref{fig:rb} sketches the outline of an RB circuit.

\begin{enumerate}
  \item Randomly select $m$ gates from the Clifford group and arrange
    them in any order.
  \item Select the Clifford gate (or gates, if performing RB on more
    than one qubit) that will reverse the operation of the entire
    preceding sequence of $m$ Clifford gates,
\begin{equation}
C_{m+1}=\left(\prod_{i=1}^{m} C_i\right)^\dagger;
\end{equation} 
execute this as the $m+1$th gate.
  \item Measure the qubit(s). If the output state is not equal to the
    input state, then an error has occurred somewhere in the whole
    circuit.
  \item Change the number $m$ and perform steps 1 to 3 again.
\end{enumerate}

By performing the above operation for various values of $m$, the
attenuation of the fidelity as a function of the circuit length can be
fitted. In this way, it is possible to estimate the average value of
errors per gate included in the gate set.  Of course, SPAM (state
preparation and measurement) errors are included in all the quantum
circuits, but this effect can be compensated for by calculating the
attenuation.

The gate error rates reported via Qiskit are derived using a procedure
similar to this, executed daily on Tokyo.  These values are not a
complete description of the behavior of the system under all
circumstances, but they are valuable and derived at reasonable cost.
One of our primary research questions, then, is whether such numbers
are \emph{good enough} to enable us to extract the maximum success
probability from the machine.

\subsection{Architecture-Aware Compilation}

Adapting to the topology of the processor using architecture-aware
circuit design and compilation has been a research topic since the
early proposals for large-scale systems and
applications~\cite{thaker06:_cqla,isailovic06:_interconnect,fowler04:_shor_implem,van-meter04:fast-modexp}.
Those early studies focused on the impact on execution time.

Recently, researchers have begun paying attention to fidelity
improvements.  For example, in the study by Zulehner \emph{et al.},
the cost function is computed by assigning a cost of 10 to a two-qubit
gate, versus a cost of one for a single qubit
gate corresponding roughly to difference in error penalty~\cite{zulehner2018mapping}.  However, in that work, no
consideration is given to the qubit-to-qubit variance in error rate.
Tannu \emph{et al.} and Finigan \emph{et al.} maximize ESP considering
errors per quantum bit~\cite{tannu2018case,finigan2018qubit}.
Finigan's research showed that ESP was improved by verifying the
optimized circuit with IBM's 16-qubit machine.

%% file: cost.tex
\section{Long-Distance CNOTs and Making Choices}
\label{sec:score}
Each $G$, $B$, and $S$ error above will have an error rate $\varepsilon$ dependent on type and location. Once assigned locations, we can compose our Estimated Success Probability for a sequence of gates to be
\begin{equation}
ESP = \prod_{i} (1 - \varepsilon_{i}). \label{eq:esp_single}
\end{equation}

To evaluate this score function, we conducted the following two experiments.
\subsection{Path selection for remote CNOT}

Most quantum algorithms use many CNOT gates.  In an architecture such
as Fig.~\ref{fig:IBMQ20}, it is not always possible to arrange the
control qubit and the target qubit close to each other. If not,
it is necessary to connect the qubits with the remote CNOT gate, or to
move via SWAP to the vicinity.  We define the following problem and
assess our ability to select the optimal solution using ESP by
experiment on Tokyo.

\begin{figure}[htbp] 
    \centering
	\includegraphics[height=3cm]{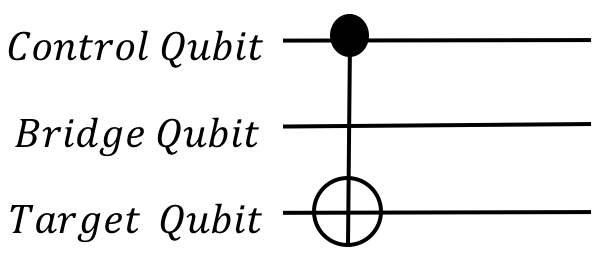} 
	\caption{Circuit A, a CNOT gate skipping across another qubit}
	\label{remoteCNOT}
\end{figure}

\begin{figure}[htbp]
\begin{minipage}[b]{0.49\hsize}
	\centering
	\includegraphics[width=2.5cm]{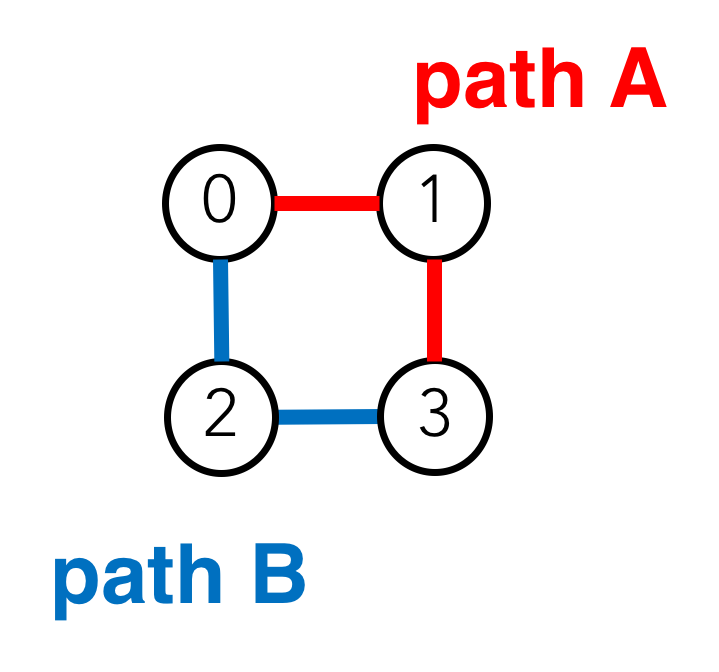}
	\subcaption{2-hop path selection}
	\label{2hoppath}
\end{minipage}%
\begin{minipage}[b]{0.49\hsize}
	\centering
	\includegraphics[width=\hsize]{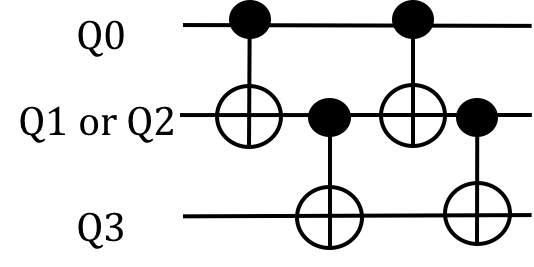}
	\subcaption{2-hop path circuit}
	\label{2hopcircuit}
\end{minipage}
\caption{(a) is two examples of CNOT path selection for 2 hops. (b) is circuit for (a). Both ends are the same qubits, and only the bridge qubit is different.}
\end{figure}

\begin{figure}[htbp]
\begin{minipage}[b]{0.49\hsize}
    \centering
	\includegraphics[width=\hsize]{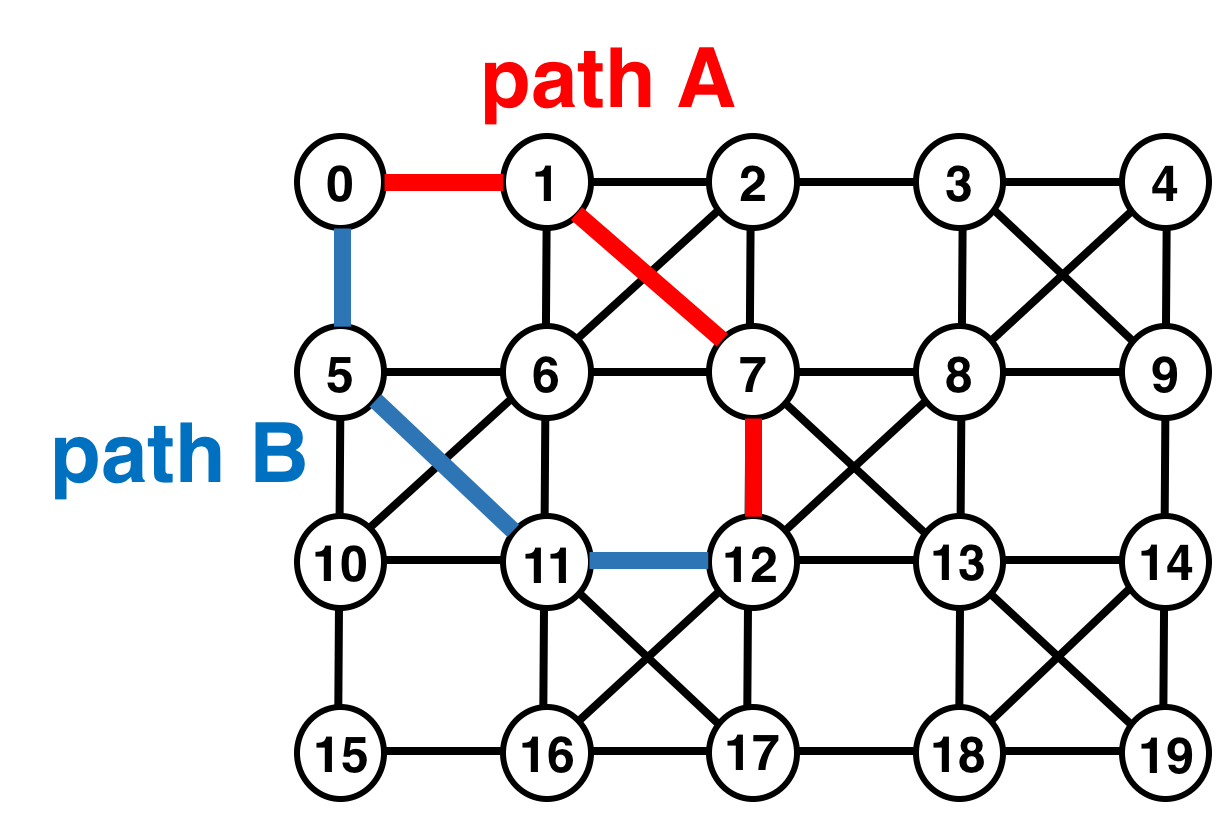}
	\subcaption{3-hop path selection}
	\label{3hoppath}
\end{minipage}%
\begin{minipage}[b]{0.49\hsize}
	\centering
	\includegraphics[width=\hsize]{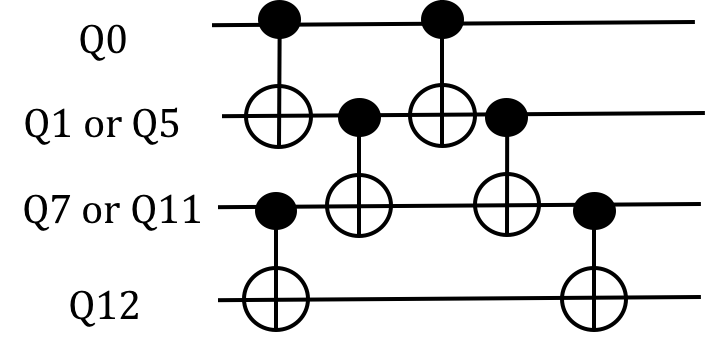}
	\subcaption{3-hop path circuit}
	\label{3hopcircuit}
\end{minipage}
\caption{As in the case of 2 hops, (a) shows two examples of 3-hop paths, and (b) shows a circuit for this path.}
\end{figure}

\begin{figure}[htbp]
	\centering
\begin{minipage}{0.8\hsize}
	\centering
	\includegraphics[height=4.5cm]{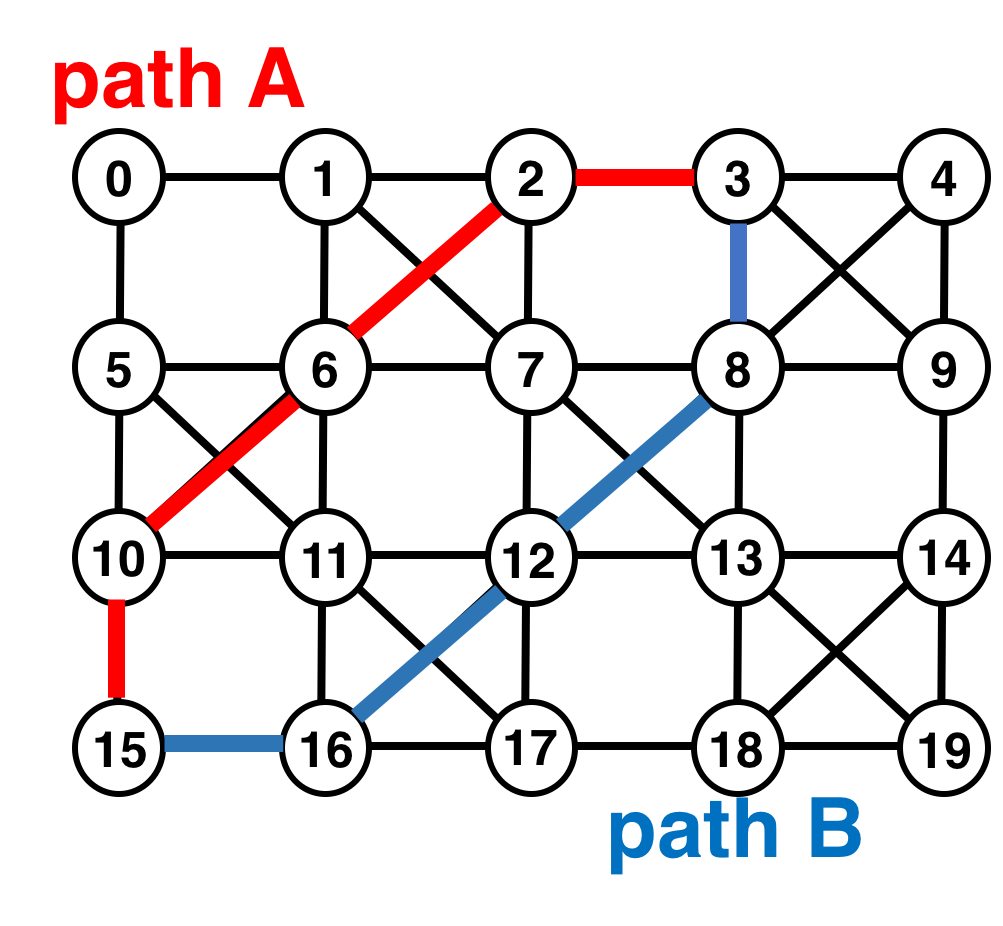}
	\subcaption{4-hop path selection}
	\label{4hoppath}
\end{minipage}
\begin{minipage}{1.0\hsize}
	\centering
	\includegraphics[width=0.9\hsize]{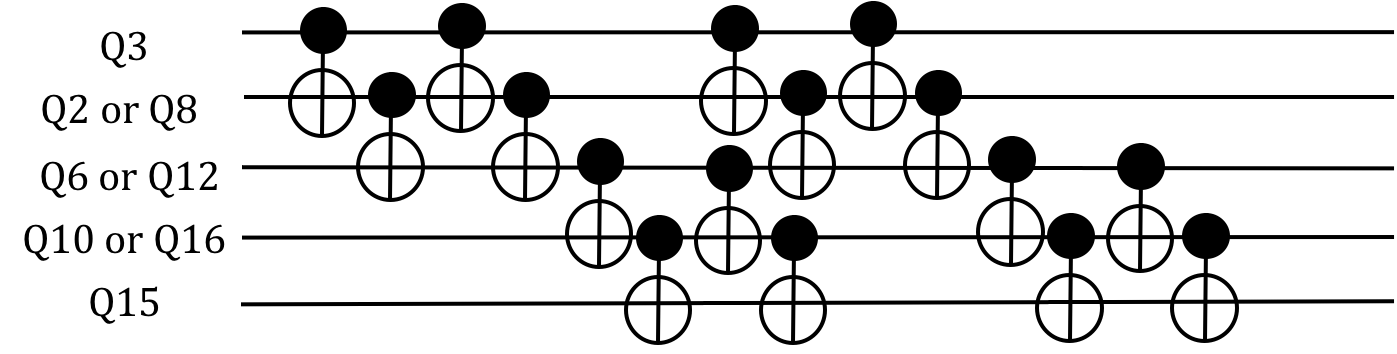}
	\subcaption{4-hop path circuit}
	\label{4hopcircuit}
\end{minipage}
	\caption{(a) shows two examples of 4-hop paths for a CNOT between Q3 and Q15, and (b) shows a circuit for this path.} 
\end{figure}

\begin{itembox}[l]{Problem 1: CNOT Path Selection}
The programmer wishes to execute the circuit shown in Fig.~\ref{remoteCNOT}.
When the starting point (control qubit) and the end point (target qubit) for the Bi-Qubit gate (CNOT) are not neighbors, which path is the highest fidelity?
\end{itembox}
We conducted the following experiments.
\begin{enumerate}
\item Select the path predicted to complete with the highest fidelity using our proposed score function (ESP). Fig.~\ref{2hoppath}, Fig.~\ref{3hoppath}, and Fig.~\ref{4hoppath} show options for 2, 3, and 4 hops path selection. The corresponding circuits are in Fig.~\ref{2hopcircuit}, Fig.~\ref{3hopcircuit}, and Fig.~\ref{4hopcircuit}.
\item Execute both paths on the actual machine for 1000 shots. For simplicity, $\ket{000}$ is used as the input state.
If the state of the target qubit is not equal to the state of the control qubit($\ket{0}$), the path can be regarded as including errors.  
As a result, the optimum path (success probability is the highest) is determined.
\item If the optimal path matches the path selected in step 1, path selection can be regarded as successful.
\end{enumerate}
As shown in Fig.~\ref{fig:IBMQ20}, the maximum eccentricity of each qubit is 4 on the largest processor made available by IBM (as of December 5, 2018). The eccentricity is the distance from a certain vertex to the furthest vertex on the same graph. Since it is sufficient to connect CNOT using routes of the number of hops, experiments were conducted up to 4 hops.\\


\begin{itembox}[l]{Result 1}
ESP was able to select the better of two routes with accuracy of 70\% at 2 hops, 66.6\% at 3 hops and 62.5\% at 4 hops.
\end{itembox}
Of course, this leaves up to 37.5\% of cases in which ESP leads us to the wrong choice. Possible reasons include:
\begin{enumerate}
\item The reported error value used to calculate ESP contains both bit-flip and phase-flip errors, but our experiments reveal only bit-flip errors.
\item Long-distance resonances and cross-talk within the system mean that a series of gates doesn't behave the same as an isolated gate. 
\item The actual state of the machine drifts faster than the calibration (Randomized Benchmarking) data is updated, so the gate error rates we use may be out of date.
\end{enumerate}
We don't believe that reason 3 is an important effect. Our future plans include tests that will distinguish the relative importance of reasons 1 and 2.

\subsection{Circuit Selection for Remote CNOT}
\begin{figure}[htbp]
\begin{minipage}{0.49\hsize}
	\centering
	\includegraphics[width=4cm]{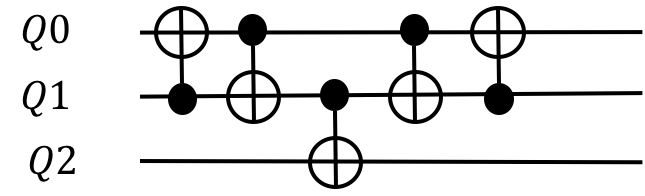}
	\subcaption{circuit B}
	\label{circuit B}
\end{minipage}%
\begin{minipage}{0.49\hsize}
	\centering
	\includegraphics[width=4cm]{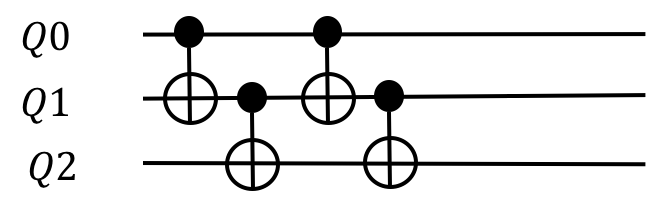}
	\subcaption{circuit C}
	\label{circuit C}
\end{minipage}
\begin{minipage}{0.49\hsize}
	\centering
	\includegraphics[width=4cm]{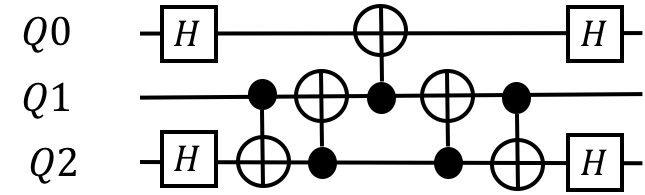}
	\subcaption{circuit D}
	\label{circuit D}
\end{minipage}
\caption{Circuits equivalent to the circuit shown in Fig.~\ref{remoteCNOT}}
\label{circuitbcd}
\end{figure}

\begin{itembox}[l]{Problem 2: Circuit Selection}
The programmer wishes to execute the circuit shown in Fig.~\ref{remoteCNOT}. When the CNOT gate is executed with remote qubits, there are multiple theoretically equivalent circuits as shown in Fig.~\ref{circuitbcd}.

However, considering errors, these are not necessarily equivalent. Which circuit has the highest fidelity?
\end{itembox}

We conducted the following experiments on Tokyo.
\begin{enumerate}
\item Select the circuit predicted to have the highest fidelity from B, C, and D in Fig.~\ref{circuitbcd} using our proposed score function (ESP).
\item Execute all three circuits on IBM Q20 Tokyo for 1000 shots. For simplicity, $\ket{000}$ is used as the input state. 

If the state of the target qubit is not equal to the state of the control qubit ($\ket{0}$), the circuit has incurred an error.

As a result, the optimum path (one with highest success probability) is determined.
\item If the optimal circuit matches the circuit selected in step 1, path selection can be regarded as successful.
\end{enumerate}

\begin{itembox}[l]{Result 2}
ESP was able to select the optimal circuit among three candidates with accuracy of 40\%. 
\end{itembox}
ESP selected Circuit B 35\% of the time, circuit C 10\% of the time,
and circuit D the remaining 55\%. Experimentally, the optimal circuit
was B for 50\%, C for 15\%, and D for 35\%. ESP is better than random,
which would select the optimal circuit only $\frac{1}{3}$ of the time.
\begin{figure}[htbp]
    \centering
	\includegraphics[width=0.8\hsize]{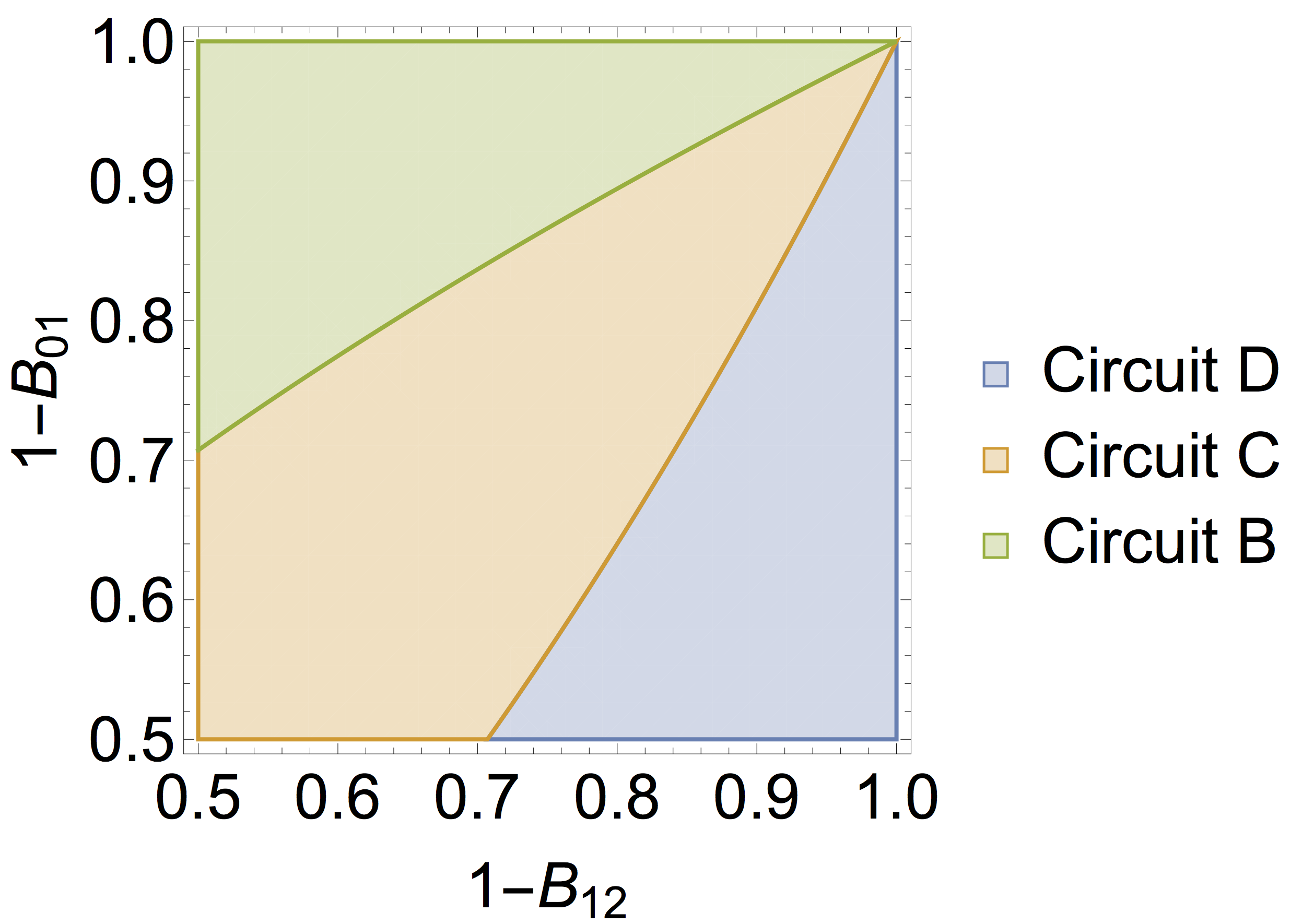}
	\caption{Estimated Best 2-hop Circuit\\
	If the error rates on the two edges are balanced, circuit C will be the best choice.
	More often on Tokyo, B or D is best.}
	\label{estimatedbest}
\end{figure}

When ESP is used, which circuit is selected based on the magnitude of Bi-Qubit gate error (B) as shown in Fig.\ref{estimatedbest}. $B_{01}$ is the Bi-Qubit gate error for Q0 and Q1. $B_{12}$ is the Bi-Qubit gate error for Q1 and Q2. Intuitively, because C is one fewer two-qubit gate, we would expect it to be the best most of the time.
In our experimental results, circuit C is chosen less often than B and
D. This is because the difference between $B_{01}$ and $B_{12}$ is
large in the current processor.

If only the number of CNOT gates is used to predict fidelity of the
quantum circuit, this will be the best choice only 15 \% of the time.

\begin{figure}[htbp]
\begin{minipage}{0.44\hsize}
	\centering
	\includegraphics[width=\hsize]{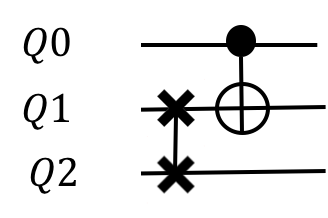}
	\subcaption{SWAP Q1, Q2}
	\label{swapa}
\end{minipage}%
\begin{minipage}{0.44\hsize}
	\centering
	\includegraphics[width=\hsize]{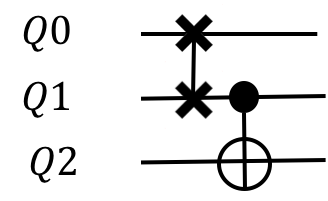}
	\subcaption{SWAP Q0, Q1}
	\label{swapb}
\end{minipage}
	\caption{Circuits including SWAP gates}
	\label{swap}
\end{figure}

\begin{figure}[htbp]
	\begin{minipage}{0.49\hsize}
	\centering
	\includegraphics[width=\hsize]{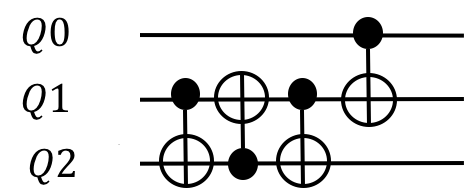}
	\subcaption{Circuit E}
	\label{circuite}
	\end{minipage}%
	\begin{minipage}{0.49\hsize}
	\centering
	\includegraphics[width=\hsize]{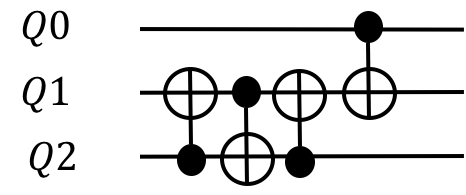}
	\subcaption{Circuit F}
	\label{circuitf}
	\end{minipage}
	\begin{minipage}{0.49\hsize}
	\centering
	\includegraphics[width=\hsize]{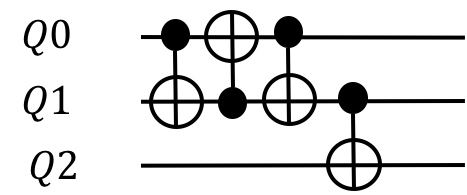}
	\subcaption{Circuit G}
	\label{circuitg}
	\end{minipage}%
	\begin{minipage}{0.49\hsize}
	\centering
	\includegraphics[width=\hsize]{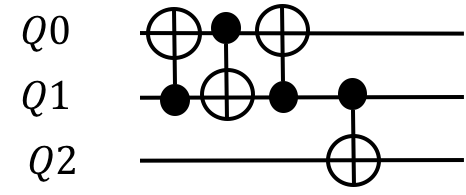}
	\subcaption{Circuit H}
	\label{circuith}
	\end{minipage}
\caption{Circuits equivalent to Fig. \protect\ref{swap}. SWAP gate can be implemented with three CNOT gates.}
\end{figure}

\begin{itembox}[l]{Problem $2'$: Circuit Selection  (including SWAP)}
  Perform the same circuit as Problem 2 with additional circuit
  options, including permitting the relocation of a qubit,
  using the SWAP gate shown in Fig.~\ref{swap}. Which circuit is the
  highest fidelity?
\end{itembox}
Because Tokyo doesn't implement a SWAP gate natively, we decompose a SWAP into
three CNOT gates. Considering the direction of CNOT gates, we have two possible decompositions.
Circuits shown in Fig.~\ref{swapa} can be
implemented into circuits shown in Fig.~\ref{circuite} and
Fig.~\ref{circuitf}. Circuits shown in Fig.~\ref{swapb} can be
implemented into circuits shown in Fig.~\ref{circuitg} and
Fig.~\ref{circuith}.  We conducted the experiments in the same way as
problem 2, using circuits B to H.

\begin{figure}[htbp]
    \centering
	\includegraphics[width=1\hsize]{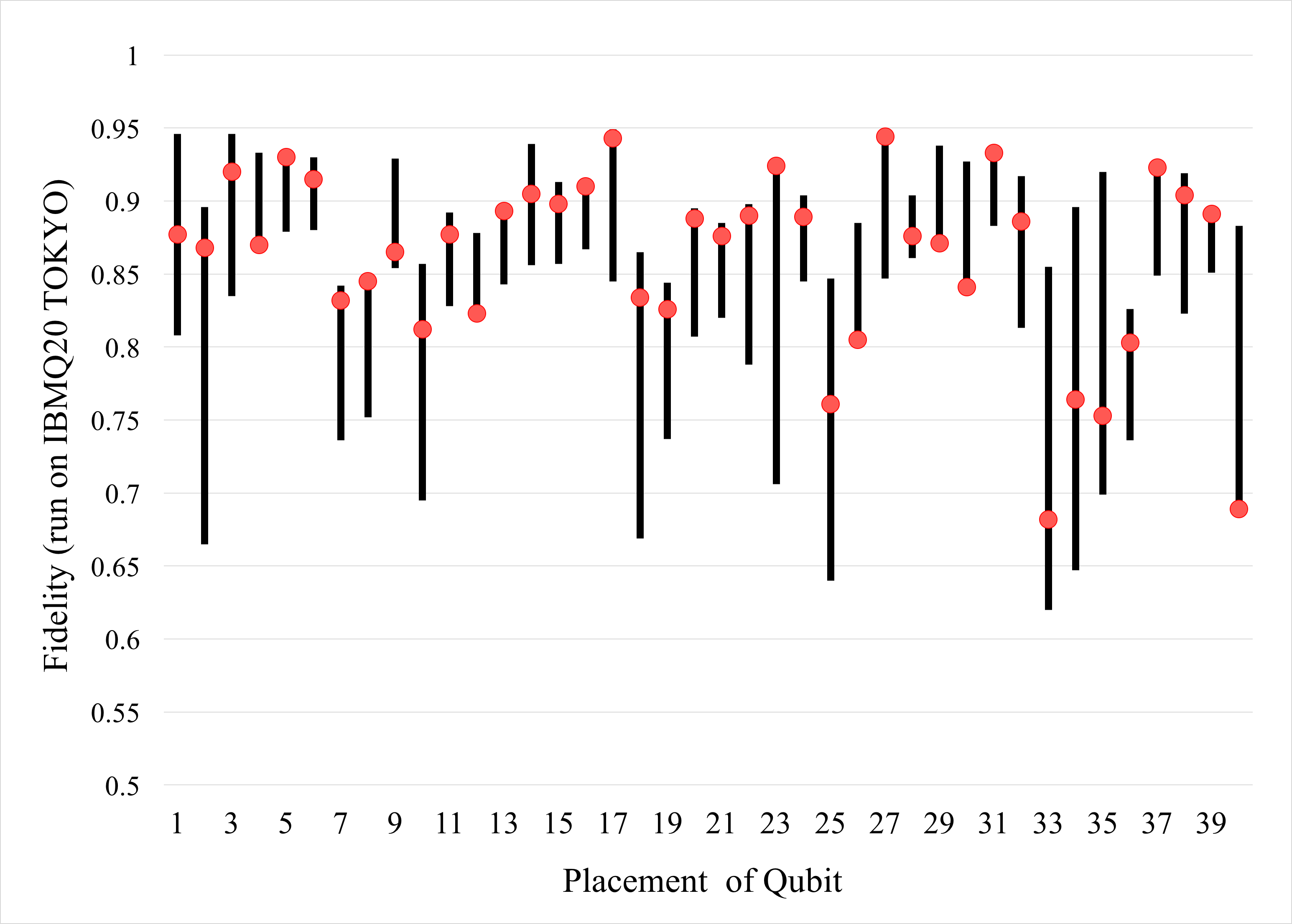}
	\caption{Fidelity of Selected Circuit \\ Seven candidate
          circuits (circuits B to H) are evaluated at forty different
          locations across the surface of Tokyo.  Red points represent
          the fidelity of the circuit selected using ESP. The top of
          the black bar shows the highest success probability achieved
          by any of the seven and the bottom of the bar shows the
          lowest one.}
	\label{fidelity_selected}
\end{figure}

\begin{itembox}[l]{Result $2'$}
  ESP was able to select the optimum circuit from among seven candidates 25\% of the time.
  Fig.\ref{fidelity_selected} shows the fidelity of selected circuit
  relative to the other candidate circuits.
\end{itembox}

The average of the ESP of the circuit selected was 0.8255. On the
other hand, the fidelity obtained by executing the circuit selected by
ESP on Tokyo was 0.8609.  With seven candidates to choose from, ESP's
25\% is better than selecting randomly, which will choose the optimal
circuit only $\frac{1}{7}$th of the time.  In most cases, we chose an
above-average circuit.

The fidelity of the selected circuit greatly exceeds ESP.  This may be
because we used only the 0 state at the input of this experiment,
whereas the single-qubit gate error used to compute ESP is for a dense
gate set. Alternatively, this may be because our experiments are more
limited than full tomography and do not reveal phase flip errors.

In order to show this, we experimented with $\ket{+++}$ input state in Problem $3'$.

\begin{itembox}[l]{Problem 3: Circuit Selection  (on Poughkeepsie)}
Perform the same circuit as Problem2’ on  the other 20-qubit processor called Poughkeepsie. Is there any change in the reliability of circuit selection by ESP?
\end{itembox}
IBM released a quantum processor called Poughkeepsie to IBMQ network members in the winter of 2018.
Table \ref{tab:specification} shows the performance specifications of Poughkeepsie.

\begin{table*}[t]
\centering
\caption{the performance specifications of two IBMQ 20-qubit systems named Tokyo and Poughkeepsie
}
\label{tab:specification}

\begin{tabular}{|r|l|l|}
\hline
\multicolumn{1}{|l|}{}                     & \begin{tabular}[c]{@{}l@{}}tokyo\\ (2nd gen 20-qubit system)\end{tabular} & \begin{tabular}[c]{@{}l@{}}poughkeepsie\\ (3rd gen 20-qubit system)\end{tabular} \\ \hline
Mean of Two-qubit (CNOT) error rates $\times 10^{-2}$ & 2.84                                                                      & 2.25                                                                             \\ \hline
best                                       & 1.47                                                                      & 1.11                                                                             \\ \hline
worst                                      & 7.12                                                                      & 6.11                                                                             \\ \hline
Mean of Single-qubit error rates $\times 10^{-3}$     & 1.99                                                                      & 1.07                                                                             \\ \hline
best                                       & 0.64                                                                      & 0.52                                                                             \\ \hline
worst                                      & 6.09                                                                      & 2.77                                                                             \\ \hline
\end{tabular}
\end{table*}

\begin{figure}[htbp]
    \centering
	\includegraphics[width=1\hsize]{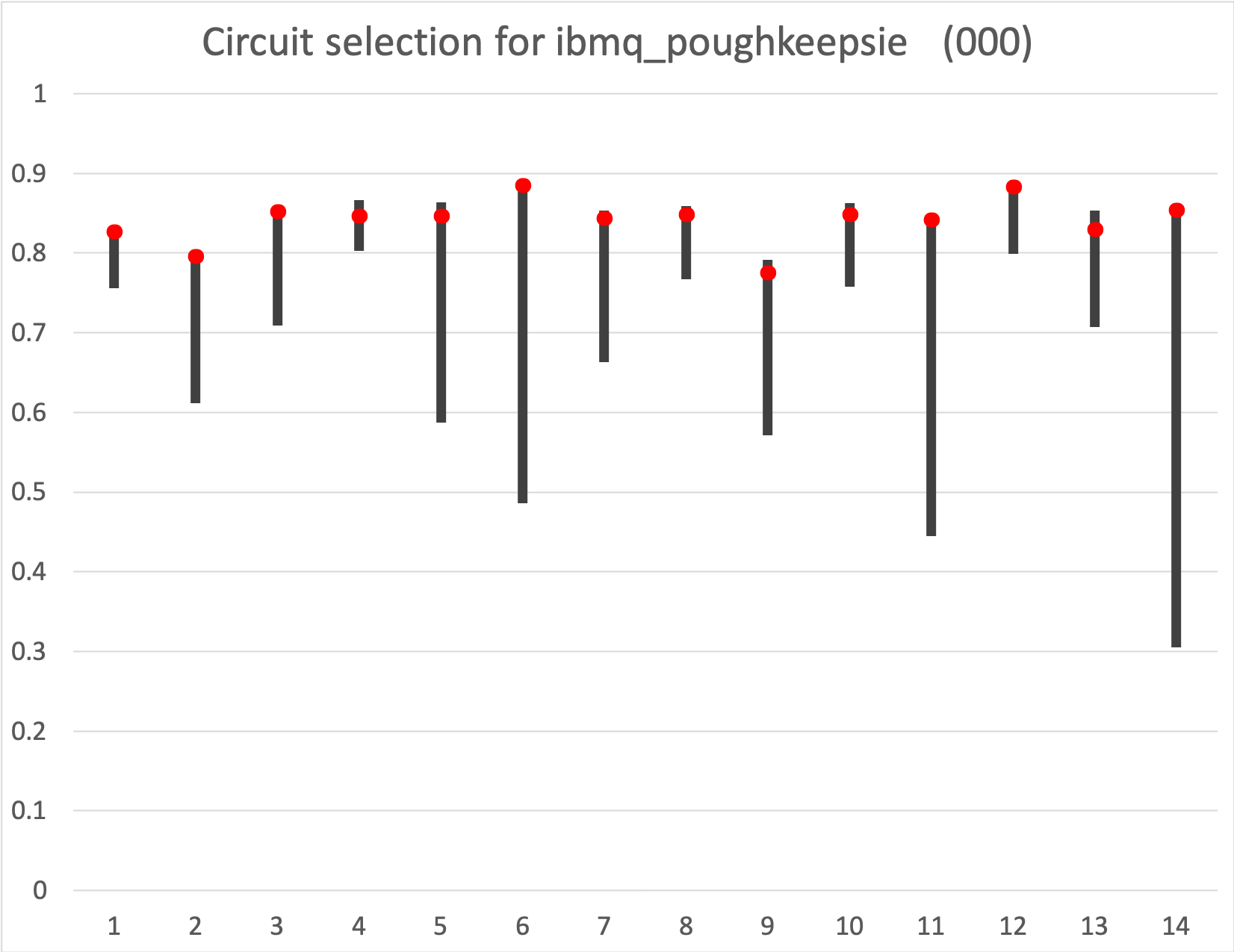}
	\caption{Fidelity of Selected Circuit on Poughkeepsie \\ 
	The experimental results were plotted in the same manner as in Fig.\ref{fidelity_selected}. The upper limit of the bar and the red dot is close.
	}
	\label{fidelity_selected_po}
\end{figure}

\begin{itembox}[l]{Result 3}
 ESP was able to select the optimum circuit from among seven candidates 43\% of the time. Fig.\ref{fidelity_selected_po} shows the fidelity of selected circuit relative to the other candidate circuits.
\end{itembox}

Among the options, the average of the Fidelity of the circuit with the best result was 0.8492. On the other hand, the average of Fidelity of the execution result of the circuit selected by ESP was 0.8405, and the difference between these was 0.87\%. This is extremely small compared to 3.5\% in tokyo, and it can be said that the reliability is improved.

The main factor is considered to be the improvement in the reliability of fitting function of the Randomized Benchmarking. That is, there is a possibility that the value of the error is less likely to fluctuate.

\begin{itembox}[l]{Problem $3'$: Circuit Selection  ($\ket{+++}$)}
Perform the same circuit as Problem 3 but with different initial input states on Poughkeepsie. Is there any change in the reliability of circuit selection by ESP?
\end{itembox}

The same circuit selection is performed by changing the quantum state before CNOT execution to a state other than 000. The  state was used for simplicity. As an alternative to the X-axis measurement, the Z measurement was performed after the H gate was performed after the CNOT. The processor used Poughkeepsie to facilitate comparison with result 3

\begin{figure}[htbp]
    \centering
	\includegraphics[width=1\hsize]{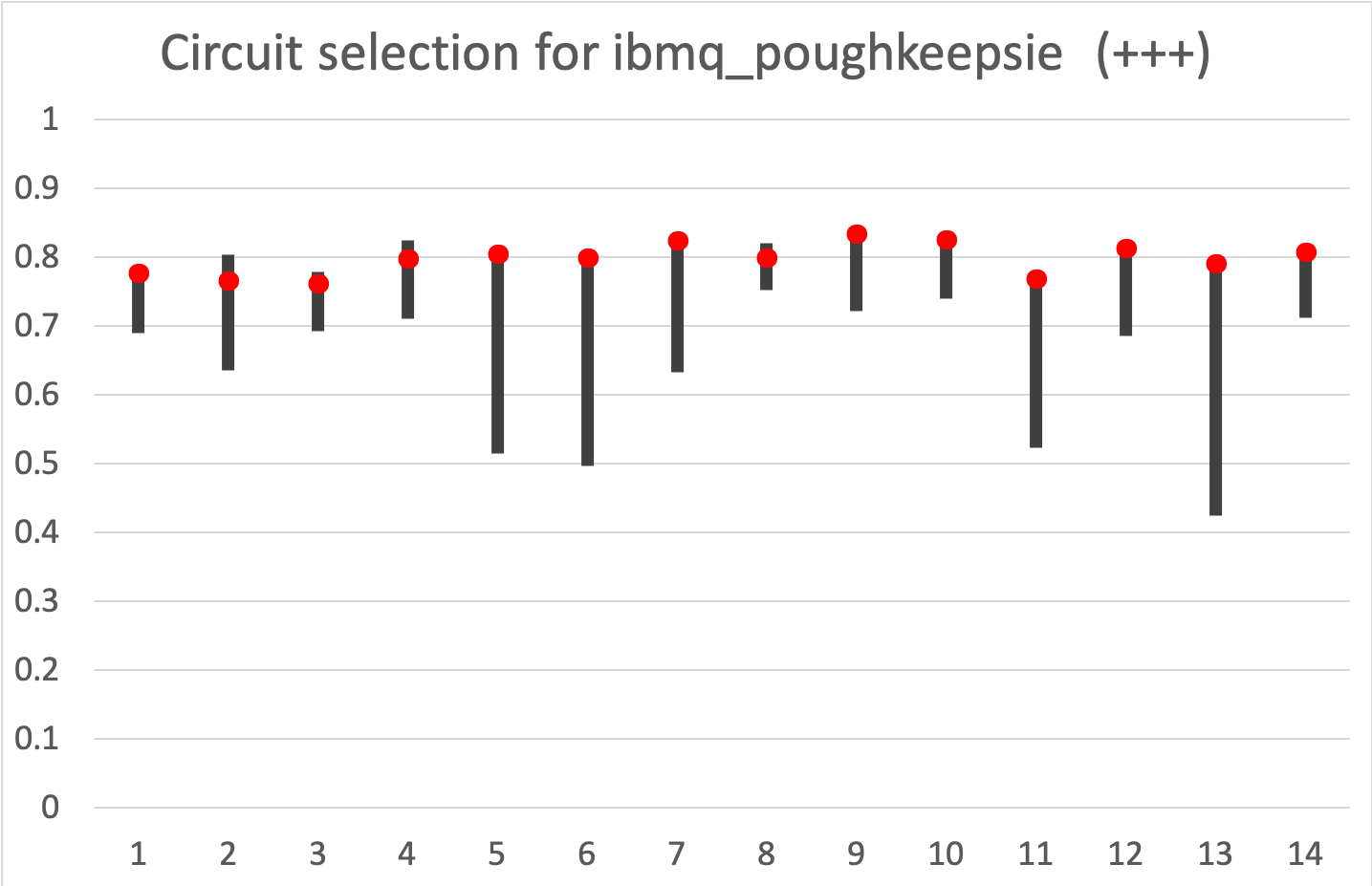}
	\caption{Fidelity of Selected Circuit on Poughkeepsie \\ 
	The experimental results were plotted in the same manner as in Fig.\ref{fidelity_selected}. The upper limit of the bar and the red dot is close.
	}
	\label{fidelity_selected_po_ppp}
\end{figure}

\begin{itembox}[l]{Result $3'$}
 ESP was able to select the optimum circuit from among seven candidates 36\% of the time. Fig.\ref{fidelity_selected_po_ppp}  shows the fidelity of selected circuit relative to the other candidate circuits.
\end{itembox}

Among the options, the average of the Fidelity of the circuit with the best result was 0.8077. On the other hand, the average of Fidelity of the execution result of the circuit selected by ESP was 0.7966, and the difference between these was 1.10\%. 

A significant improvement in accuracy is seen over result 2 ', but a little less accurate than result 3. Also, the overall fidelity itself is low.

Since the number of gates is increasing, it is estimated that the prediction accuracy has fallen.

%% file: algorithm.tex
\section{Compiling Complete Programs}
\label{sec:qopter}

In the previous section, we discuss several possible realizations of a single CNOT
gate on IBM's QX architecture and compare estimates and observational results of their probability of success.  A general purpose compiler,
however, is required to combine such realizations
in order to deal with much more complex quantum circuits with many CNOT gates.
Since there are several possible combinations of realizing the circuit,
the compiler must be able to distinguish more reliable realizations from the others.
To give a metric of several realizations, we extend the definition of Eq.~\ref{eq:esp_single}'s
Estimated Success Probability; we define the Estimated Success Probability of a
circuit $C$ as the product of ESPs of its components:
\begin{align}
    ESP(C) = \prod_{g \in C} ESP(g) = \prod_{g \in C} (1 - \varepsilon_g). \label{eq:esp_circuit}
\end{align}

In this model, we assumed that
\begin{enumerate}
    \item each gate and measurement in a circuit either succeeds completely or fails to execute and stops the whole execution; and
    \item the probabilities of such failures are independent of each other and depend only on the physical qubits on which the operations act
\end{enumerate}

With this formalization, the compiler's task is defined to be the maximization of
$ESP(C)$, and we apply a combinatorial optimization method described in the
following subsections to this end.

Although Equation \ref{eq:esp_circuit} is simple and easy to
calculate for the optimization process, we cannot use it when evaluating
optimized circuits on real NISQ computers.  The problem is that, in general, the output of a
quantum circuit is not a single value but a probability distribution.
Imagine a quantum circuit with Hadamard gates and measurements for each qubit.
Since this circuit produces a uniform distribution over
all of the possible measurements, we cannot figure out whether a single-shot measurement of
the circuit on a quantum device is a successful execution or not.  Instead, we
have to sample multiple shots from the circuit and compare the resulting empirical
distribution against the ideal distribution a noiseless quantum
computer would produce.  We propose using the KL divergence between the empirical and
ideal distributions as a distribution-based measure of a compiled circuit
and studied the relationship between ESP and KL divergence on IBM Q20 Tokyo
machine.

\subsection{Search Space}
In this section, we scrutinize the search space of the ESP maximization problem.

Each state we visit in the optimization process is parameterized by two parts: the
execution state of gates and the qubit mapping.  A gate execution state
describes which gates have already been executed and which have not.  Let $N$ be the
number of gates, then the number of execution states is $2^N$.  Moreover, for
each gate execution state, there exists
$_QP_V = Q\times (Q-1) \times \dots \times (Q-V+1)$ arrangements of quantum
variables onto the physical qubits~\cite{1810.08291}.

The search space of this problem can be visualized as a collection of hypercubes.
Each hypercube consists of a complete
set of execution states, and there is one corresponding hypercube for each
qubit mapping.  There are two types of transitions between the states: gate
execution and swap insertion.

Execution of a gate sets an execution flag of the gate execution state, moving
the state upward from the $000\dots 0$ vertex toward $111\dots 1$ in the same hypercube.  On the other hand, insertion of SWAP
gates does not change the execution state while it changes the qubit mapping.  Thus
SWAP insertion moves the state laterally, making a jump to another hypercube.
Under this setting, each path from a state at the bottom to
a state at the top constitutes a possible compiled circuit, and the compiler's task can be
interpreted as finding the best path from the bottom to the top.

As discussed in the prior sections, CNOT gates can be executed if and only if
the control qubit and the target qubit are adjacent to each other in the
physical topology.  This adjacency constraint invalidates some edges.
Moreover, to preserve the logical function of the given quantum circuit during
the compilation, some gates must be executed before other gates.  We call this
constraint of order a gate dependency.  Gate dependencies forbid some
transition edges in hypercubes. 

A naive brute-force algorithm will traverse $2^N \times {_QP_V}$ states in the
worst case, which is infeasible. Additionally,
~\cite{Siraichi:2018:QA:3179541.3168822} shows that finding the optimal path is an
NP-hard problem when the optimization target is the number of inserted SWAP
gates.  Thus, finding the optimal solution would be intractable in practice.

\subsection{The Optimization Algorithm}
\label{sec:compile-algorithm}

Therefore, we propose a beam-search based heuristic optimization algorithm.
Beam search is a modified version of breadth-first search where at each depth,
instead of adopting all candidates as breadth-first search does,
only a fixed number of promising states are searched. The number of candidates kept is called the beam
width $B$. As the beam width grows, the number of states that the algorithm visits
increases and better solutions will be found, at the cost of higher time and
space complexity. Specifically, if $B = 1$, beam search is
identical to greedy search, and to breadth-first search if $B \to \infty$.

In addition, the choice of initial qubit mappings is crucial because it corresponds to
the initial states of the beam search.  One possible way to generate initial
mappings is to assign quantum variables to physical qubits randomly.  Besides
that, several papers~\cite{Siraichi:2018:QA:3179541.3168822,1810.08291} have proposed heuristic methods to generate
proper mappings in order to get shorter or more shallow circuits.  Our approach
combines these methods.  In other words, we start the search with an initial
qubit mapping computed by a heuristic method in addition to random initial mappings.
This approach enjoys the performance of heuristics while also exploring the
possibility of nicer configurations through chance.  We denote
the number of random initial mappings as $M$.

We show the compilation algorithm in Algorithm \ref{alg:compile}.  In this
algorithm, the number of executed gates inside all the states in the next state
set ($S_{i+1}$) is incremented by one from those of the previous state set ($S_i$).
Therefore, after $N$ iterations of the outermost loop, $S_N$ contains
states with all gates executed.

Inside the loop, we iterate the state set $S_i$ and update it by the new set
$S_{i+1}$.  Since the innermost loop only checks $N$ gates, the inside of the loop runs
$N|S_i|$ times for each $i$.  Inside the loop, the dependency check
(line 7) can be done in constant time with auxiliary information
encoded in $state$. Moreover, the BEST\_SWAP function call (line 9) is also a
constant time operation with caching.  Therefore, the most compute-intensive
part of the algorithm is calculating the score of the state at line 12.
Since UPDATE\_SCORE takes $O(N)$ time (described in Section
\ref{sec:score-function}), the complexity of the update is $O(N^2|S_i|)$ for each
$i$.  On the first iteration, $|S_0| = M$ because $S_0$ is filled with initial states and
when $i > 0$, $|S_i| \leq B$ holds because beam search prunes the states.
Therefore, the total complexity of this algorithm is $O(N^2 M + N^3 B)$, where
the first term corresponds to the first iteration and the second term
to the rest.

We can ignore the complexity of pruning (line 17) as pruning takes
$O(|S_{i+1}|) = O(N|S_i|)$ time with the Floyd-Rivest algorithm
~\cite{Floyd:1975:AAS:360680.360694}, which has lower complexity than
computing the score of each new state.

\begin{figure}
\begin{minipage}{\linewidth}
\begin{algorithm}[H]
\caption{Compilation Algorithm}
\label{alg:compile}
\begin{algorithmic}[1]
    \Function{compile}{$gates, topology$}
        \State $N \leftarrow |gates|$
        \State $S_0 \leftarrow$ initialize states with INITIAL\_MAPPING($gates, topology$)
        \For {$i = 0$ upto $N$ (exclusive)}
            \State $S_{i+1} \leftarrow \{\}$
            \ForAll {$state \in S_i$}
                \ForAll {$g \in gates$}
                    \If {dependency of $g$ is satisfied}
                        \State $qubits \leftarrow state.mapping[g.qubits]$
                        \State $swap \leftarrow$ BEST\_SWAP$(qubits, topology)$
                        \State $state' \leftarrow $ $state$ with $swap$ inserted and $g$ executed
                        \State $state'.esp \leftarrow state.esp \times ESP(swap) \times ESP(g)$
                        \State UPDATE\_SCORE($state', gates, topology$)
                        \State $S_{i+1} \leftarrow S_{i+1} \cup \{state'\}$
                    \EndIf
                \EndFor
            \EndFor
            \State $S_{i+1} \leftarrow$ top-$B$ states of $S_{i+1}$
        \EndFor
        \State \Return the best $state \in S_N$
    \EndFunction
\end{algorithmic}
\end{algorithm}
\end{minipage}%
\end{figure}

\subsection{Subroutines for Compilation Algorithm}
In the following sections, we describe the three subroutines which appear in the
compilation algorithm.

\subsubsection{Scoring States}
\label{sec:score-function}
The function UPDATE\_SCORE computes the score of states. Pruning removes
states with lower scores calculated by this function.

The simplest score function is just using the ESP of executed gates.
Incorporating information of future gates can further improve the
selection of states.  In this compiler, we multiply current ESP with the
imaginary ESPs of each gate to be executed.  The imaginary ESP is the
Estimated Success Probability of SWAP gates to satisfy the adjacency constraint
under the current mapping times the success probability of the
gate. Imaginary ESP can differ from actual ESP because the current mapping
can differ from
mappings at the time of gate execution due to the insertions of SWAP gates
required for the execution of former gates.

We show this computation in Algorithm \ref{alg:update_score}.  This algorithm
enumerates unexecuted gates, so the complexity of computing the score is
$O(N)$.

\begin{figure}
\begin{minipage}{\linewidth}
\begin{algorithm}[H]
\caption{Update Score of a State}
\label{alg:update_score}
\begin{algorithmic}[1]
    \Function{update\_score}{$state, gates, topology$}
        \State $score \leftarrow state.esp$
        \ForAll {$g \in gates$}
            \If {$g$ has not been executed}
                \State $swap \leftarrow$ BEST\_SWAP$(state.mapping, g, topology)$
                \State $score \leftarrow score \times ESP(swap) \times ESP(g)$
            \EndIf
        \EndFor
        \State $state'.score \leftarrow score$
    \EndFunction
\end{algorithmic}
\end{algorithm}
\end{minipage}%
\end{figure}

\subsubsection{Heuristic Initial Mapping: Greatest Connecting Edge Mapping}
As mentioned above, our compiler adopts random and heuristic initial mappings as
the starting points of search.  The core idea of our initial heuristic
mapping is to map more significant edges between the quantum variables to
less noisy edges in the physical topology.  This idea comes from the fact
that error rates of CNOT gates are a magnitude higher than those of
single qubit gates on the IBM QX architecture~\cite{ibm:qexp}.

This heuristic starts with counting the number of CNOT gates executed over for
each pair of quantum variables.  These numbers constitute a guest graph,
a graph whose vertices correspond to each quantum variable and edges to the
number of CNOT gates over them.

Next, we scan the edges of the guest graph in a similar manner to Prim's algorithm~\cite{Prim}. However, instead
of scanning from the lowest to greatest, this algorithm scans from the greatest
to lowest, constructing a maximum spanning tree over the guest graph.

During the first scan, both endpoints of the chosen edge are not mapped yet.
Thus, we pick the least noisy edge from the physical topology and map those two variables
to this edge.

In the following scans, one endpoint of the chosen edge is already
mapped, and the other is not, so the chosen edge is the greatest edge
connecting the set of mapped variables and that of unmapped variables.  The
name of the heuristics comes from this.  Since one of the variables is already
mapped, it would be a good idea to map the other to the adjacent qubit in the
physical topology.  So we search for the free qubit adjacent to the qubit the endpoint
is mapped to and adopt the qubit with the lowest CNOT error rate between the
qubits.  We show pseudo-code in Algorithm \ref{alg:initial-mapping}.
During the scan, a mapped qubit may not have free qubits adjacent to
itself.  In that case, this algorithm skips that edge and maps unmapped
variables to the remaining qubits randomly at the end.

\begin{figure}
\begin{minipage}{\linewidth}
\begin{algorithm}[H]
\caption{Greatest Connecting Edge Mapping}
\label{alg:initial-mapping}
\begin{algorithmic}[1]
    \Function{initial\_mapping}{$gates, topology$}
        \State extract guest graph $G_V = (V_V, E_V)$ from $gates$
        \State $Mapped = \{\}$
        
        \State $(v_1, v_2) \leftarrow$ greatest edge $ \in E_V$
        \State $(q_1, q_2) \leftarrow$ lowest edge $\in topology$

        \State $M \leftarrow \{ v_1 \mapsto q_1, v_2 \mapsto q_2 \}$
        \State $Mapped \leftarrow Mapped \cup \{v_1, v_2\}$
        \State $E_V \leftarrow E_V \setminus \{(v_1, v_2)\}$

        \While {not all endpoints in $E_V$ are mapped}
            \State pick the greatest edge $(v, v') \in E_V$ where $v \in S$ and $v' \notin S$
            \If {there is a free physical qubit adjacent to $M[v]$}
                \State pick the lowest edge $(q, q') \in topology$ where $q = M[v]$ and $q'$ is a free qubit
                \State $M \leftarrow M \cup \{v' \mapsto q'\}$
                \State $S \leftarrow S \cup \{v'\}$
            \EndIf
            \State $E_V \leftarrow E_V \setminus \{(v, v')\}$
        \EndWhile

        \ForAll {$v \in V_V \setminus Mapped$}
            \State pick a random free physical qubit $q \in topology$
            \State $M \leftarrow M \cup \{v \mapsto q\}$
        \EndFor

        \State \Return M
    \EndFunction
\end{algorithmic}
\end{algorithm}
\end{minipage}%
\end{figure}

\subsubsection{Finding Best Swap Sequence}
In our compiler, we consider only the realizations of CNOT gates via the
insertion of a sequence of SWAP gates.  The compiler needs to know how to
insert a SWAP sequence to change the mapping such that the adjacency constraint is
satisfied in the new mapping.  BEST\_SWAP function computes such SWAP sequences
for each physical qubit in the topology.

The algorithm is shown in Algorithm \ref{alg:best-swap}.  The main idea of this
algorithm is to find the best ``meeting edge'' for the given pair of qubits.
Each qubit will be moved to the endpoints of the meeting edge, and finally, we
execute the CNOT gate over the edge.  The ESP of this movement plus the CNOT
gate can be calculated as the product of ESP of the SWAP sequences and ESP
of the CNOT gate over the meeting edge.  We can run the shortest path algorithm
over the physical topology to find the SWAP sequence with optimal ESP.

In Algorithm \ref{alg:best-swap}, we run the shortest path algorithm in the
loop; in the implementation, however, we ran the Warshall-Floyd algorithm~\cite{Warshall:1962:TBM:321105.321107,Floyd:1962:A9S:367766.368168} at the
beginning of the compilation to compute the optimal path for all pairs of
qubits, and we fetch the result.  Moreover, the results of calls to BEST\_SWAP
themselves can be cached.  Therefore, we compute BEST\_SWAP for all possible
combinations of qubits before Algorithm \ref{alg:compile} and we use the cached
sequence in the main loop.

\begin{figure}
\begin{minipage}{\linewidth}
\begin{algorithm}[H]
\caption{Find The Best Swap Sequence}
\label{alg:best-swap}
\begin{algorithmic}[1]
    \Function{best\_swap}{$qubits, topology$}
        \If {$qubits.len() == 1$}
            \State \Return []
        \EndIf

        \State $ESP_{max} \leftarrow 0$
        \State $swap_{max} \leftarrow []$
        \ForAll {$(q_0, q_1) \in topology$}
            \State $swap_0 \leftarrow$ find the swap sequence between $qubits[0]$ and $q_0$ with greatest $ESP(swap_0)$
            \State $swap_1 \leftarrow$ find the swap sequence between $qubits[1]$ and $q_1$ with greatest $ESP(swap_1)$
            \State $ESP \leftarrow ESP(swap_0) \times ESP(swap_1) \times ESP(\text{CNOT over } q_0 \text{ and } q_1)$
            \If {$ESP > ESP_{max}$}
                \State $ESP_{max} \leftarrow ESP$
                \State $swap_{max} \leftarrow swap_0 + swap_1$
            \EndIf
        \EndFor

        \State \Return $swap_{max}$
    \EndFunction
\end{algorithmic}
\end{algorithm}
\end{minipage}%
\end{figure}

\subsection{Experimental Evaluation of Compilation}

\subsubsection{Probability Distribution Based Evaluation}

In this section, we explain the experimental evaluation of the compilation on
IBM's Q20 Tokyo machine.  As described in Equation \ref{eq:esp_circuit}, our
compiler uses a simple multiplicative error model to optimize the ESP of
quantum circuits.  However, ESP has difficulty as an metric of the quality
of the compiled circuits when it comes to the evaluation of real quantum
hardware.

The difficulty is that the experimental result of a quantum circuit cannot be
judged either successful or unsuccessful in general.  Imagine a quantum circuit
with a Hadamard gate and measurement for each qubit.  This circuit will produce
a uniform distribution over the all possible measurements.  Therefore, we
cannot decide whether the circuit succeeded or failed from one shot of measurement.
Instead, we have to focus on how close the empirical probability distribution
sampled by running the compiled circuit on a NISQ machine multiple times and the
ideal distribution an imaginary noiseless quantum computer will produce are.

To deal with this difficulty, we propose a probability distribution based
measure of compiled circuits.  For each quantum circuit $C$, let $P_{ideal}$ be
the ideal distribution of circuit $C$ and $P_{empirical}$ be the empirical
distribution we observed by running compiled circuit on a machine.  Then, we
define KL divergence $D_{KL}(P_{ideal} || P_{empirical})$ between those two
probability distributions as follows:
\begin{align}
    D_{KL}(P_{ideal} || P_{empirical}) = \sum_{x} P_{ideal}(x) \log \frac{P_{ideal}(x)}{P_{empirical}(x)}
\end{align}
where $x$ runs over the possible measurement results. It's known that
$P_{ideal} = P_{empirical} \,a.e. \Leftrightarrow D_{KL}(P_{ideal} || P_{empirical}) = 0$
so the lower $D_{KL}$ is, the better the quality of the compiled circuit is.

Boixo \emph{et al.} proposed the cross entropy, which is equal to KL divergence plus a certain
offset, as a benchmark for verifying Quantum Supremacy~\cite{1203.5813} in NISQ
devices~\cite{2018NatPh..14..595B}, however, no papers yet use such
measures for compiling quantum circuits as far as we know.

\subsubsection{Experiment and Result}

\begin{table*}[t]
    \centering
    \caption{
    The number of gates and compilation time for each adder.
    $Q_{input}$ denotes the number of input qubits of the adders.
    $g_{ori}$ is the number of CNOT and single qubit gates in the pre-compiled circuit.
    $g_{min}, g_{median}, g_{max}$ shows the minimum, median, and max number of gates in the compiled circuits, respectively.
    $T_{min}, T_{median}, T_{max}$ are the minimum, median and max time in seconds needed to compile the circuit, respectively.
    We ran the compiler with Intel Core i7-8550U and 16GB RAM.
    The compiler is implemented in Rust and consists of 1894 lines of code.
    }
    \label{tab:compilation}
    \begin{tabular}{|c|c|c|c|c|c|c|c|c|c|c|c|c|c|}
        \hline
        & & \multicolumn{6}{|c|}{beam search} & \multicolumn{6}{|c|}{random selection} \\
        \hline
        $Q_{input}$ & $g_{ori}$ & $g_{min}$ & $g_{median}$ & $g_{max}$ & $T_{min}$ & $T_{median}$ & $T_{max}$ & $g_{min}$ & $g_{median}$ & $g_{max}$ & $T_{min}$ & $T_{median}$ & $T_{max}$ \\
        \hline
        1 & 45  & 48  & 66  & 81  & 2.96 & 3.11 & 3.34 & 69  & 96  & 129 & 0.022 & 0.032 & 0.040 \\ \hline
        2 & 82  & 100 & 121 & 145 & 9.47 & 10.2 & 11.6 & 103 & 127 & 211 & 0.032 & 0.042 & 0.053 \\ \hline
        4 & 156 & 255 & 317 & 390 & 31.3 & 33.6 & 36.1 & 239 & 290 & 380 & 0.051 & 0.063 & 0.131 \\ \hline
    \end{tabular}
\end{table*}

\begin{figure}
    \input{adder.tex}
    \caption{
    Cucarro's adder circuit of 2-qubit inputs and carry-in ($\vert c \rangle$)
    and carry-out ($\lvert z \rangle$). MAJ (MAJority) gate consists of one
    Toffoli gate and two CNOT gates and computes carry bit by determining the
    majority of inputs are $\lvert 1 \rangle$ state. UMA (UnMajority and Add) is
    also made of one Toffoli and two CNOTs and computes the addition of this
    digit on $\lvert b \rangle$, uncomputing the other qubits. We denote the
    $i$'th bit of the addition as $\lvert s_i \rangle$ in the figure. Toffoli
    gates are made of 6 CNOT gates and 9 single qubit gates. Since we give
    $\lvert 0 \rangle$ for $\lvert c \rangle$ and $\lvert z \rangle$, the
    measurement of this circuit gives you the distribution of the added value.
    }
    \label{fig:adder}
\end{figure}
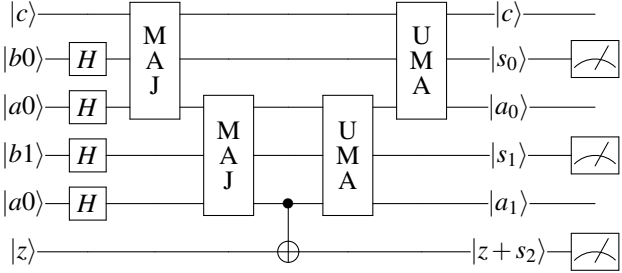
\begin{figure}
    \includegraphics[width=\hsize]{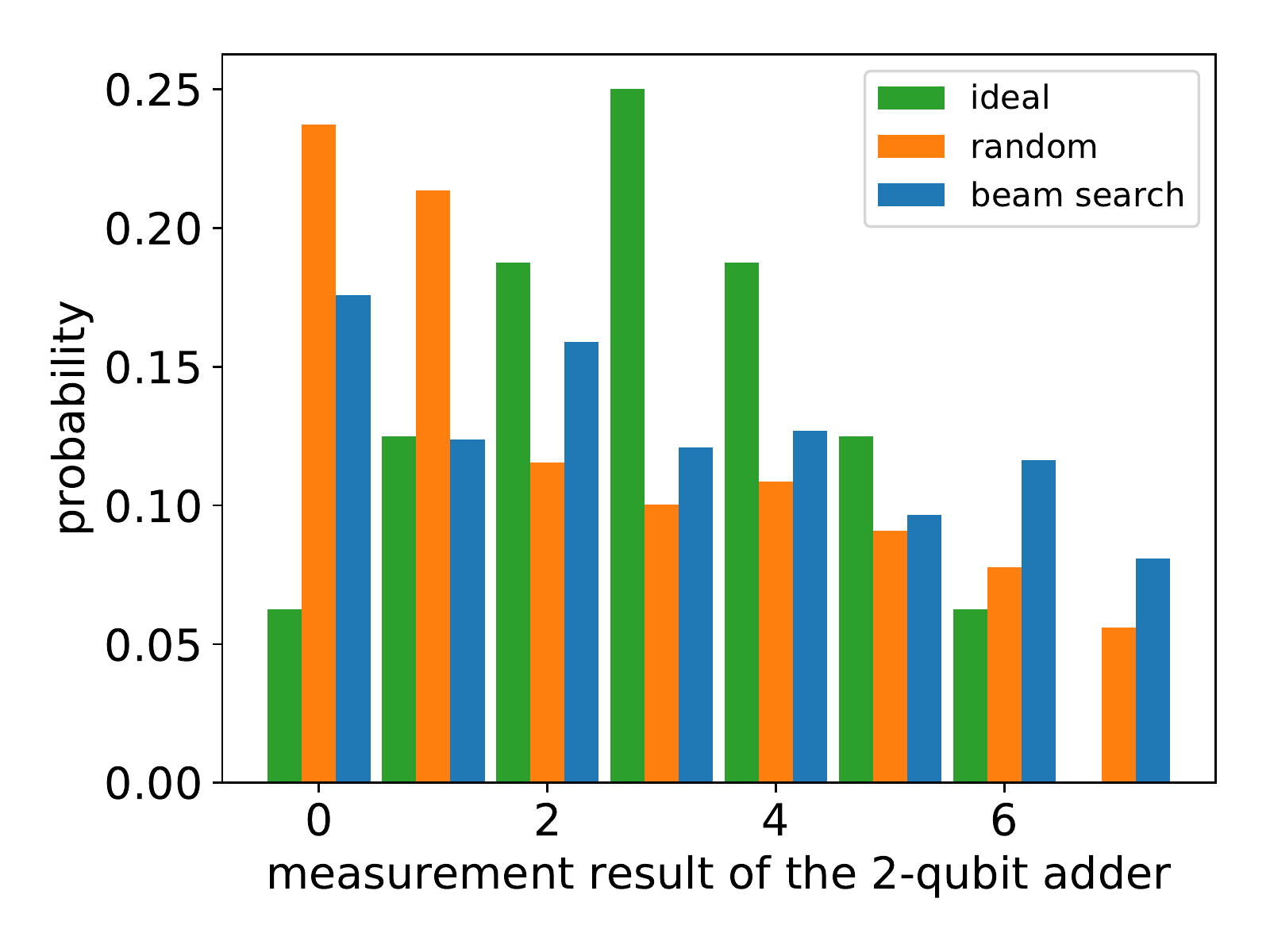}
    \caption{
    Example distributions of 2-qubit adder experiments. The green bars
    correspond to the ideal distribution calculated by classical simulation, and
    the other bars (orange and blue) show the distribution we retrieved by
    running on IBM Q20 Tokyo the circuits compiled by random selection compilation
    algorithm and beam search, respectively. In this experiment, we got
    $ESP = 0.169$ and $D_{KL} = 0.297$ for the random selection compiler, $ESP = 0.091$ and
    $D_{KL} = 0.216$ for the beam search compiler.
    }
    \label{fig:pyramid}
\end{figure}

%
%
\begin{figure}[t]
    \begin{minipage}{0.9\hsize}
        \centering
        \includegraphics[width=\hsize]{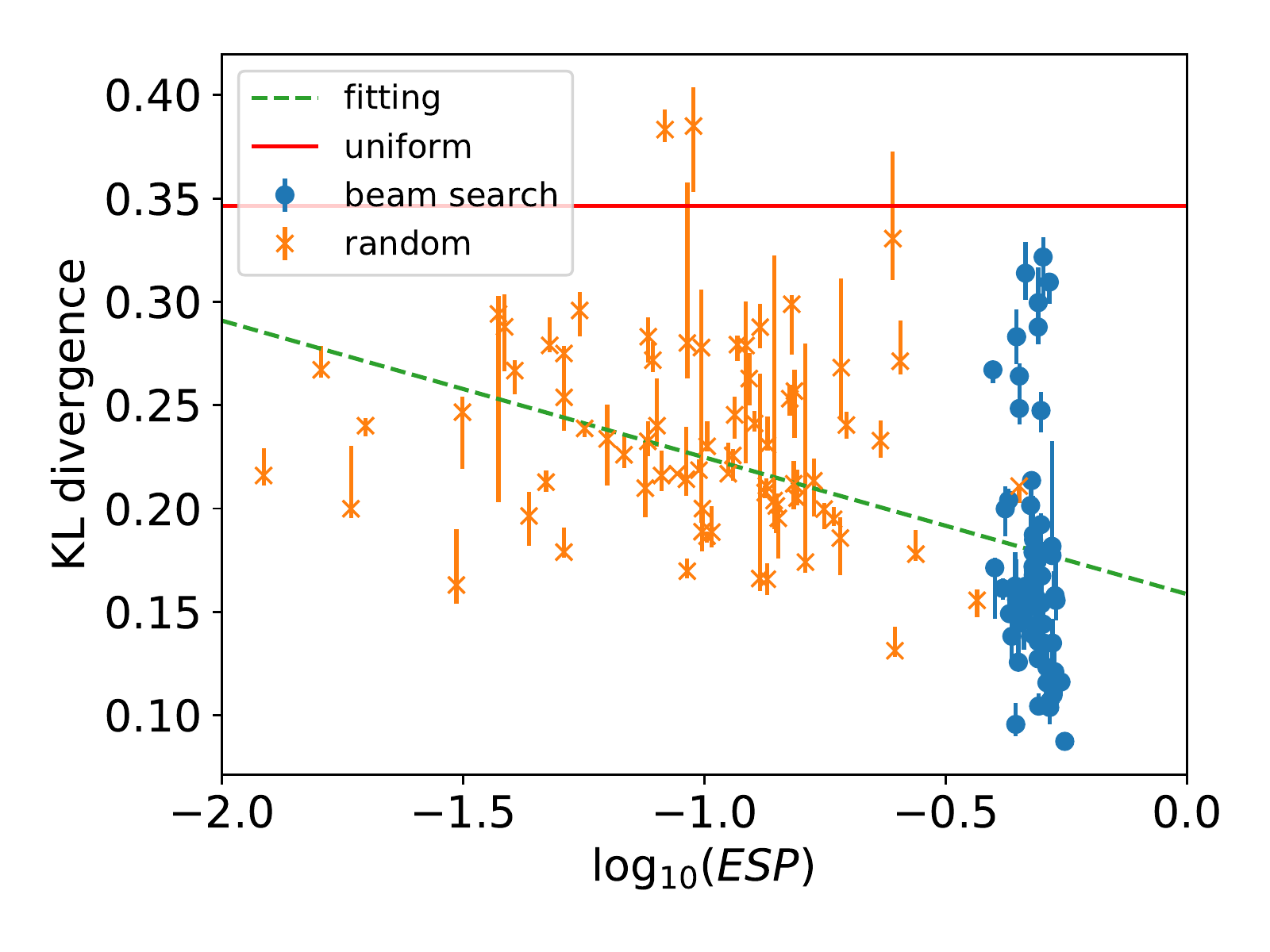}
        \subcaption{
            1-qubit adder
        }
        \label{fig:1-qubit-adder}
    \end{minipage}

    \begin{minipage}{0.9\hsize}
        \centering
        \includegraphics[width=\hsize]{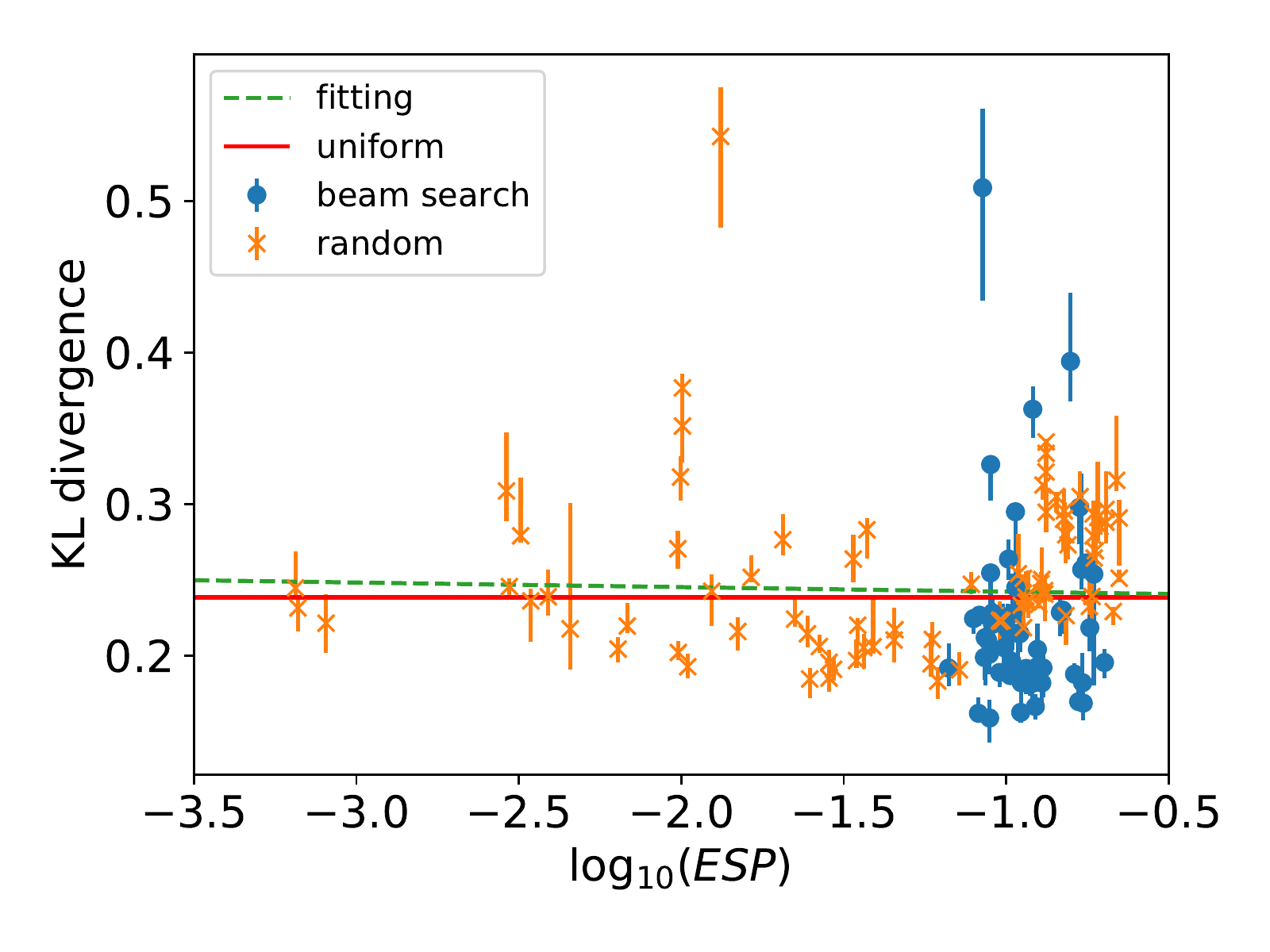}
        \subcaption{
            2-qubit adder
        }
        \label{fig:2-qubit-adder}
    \end{minipage}

    \begin{minipage}{0.9\hsize}
        \centering
        \includegraphics[width=\hsize]{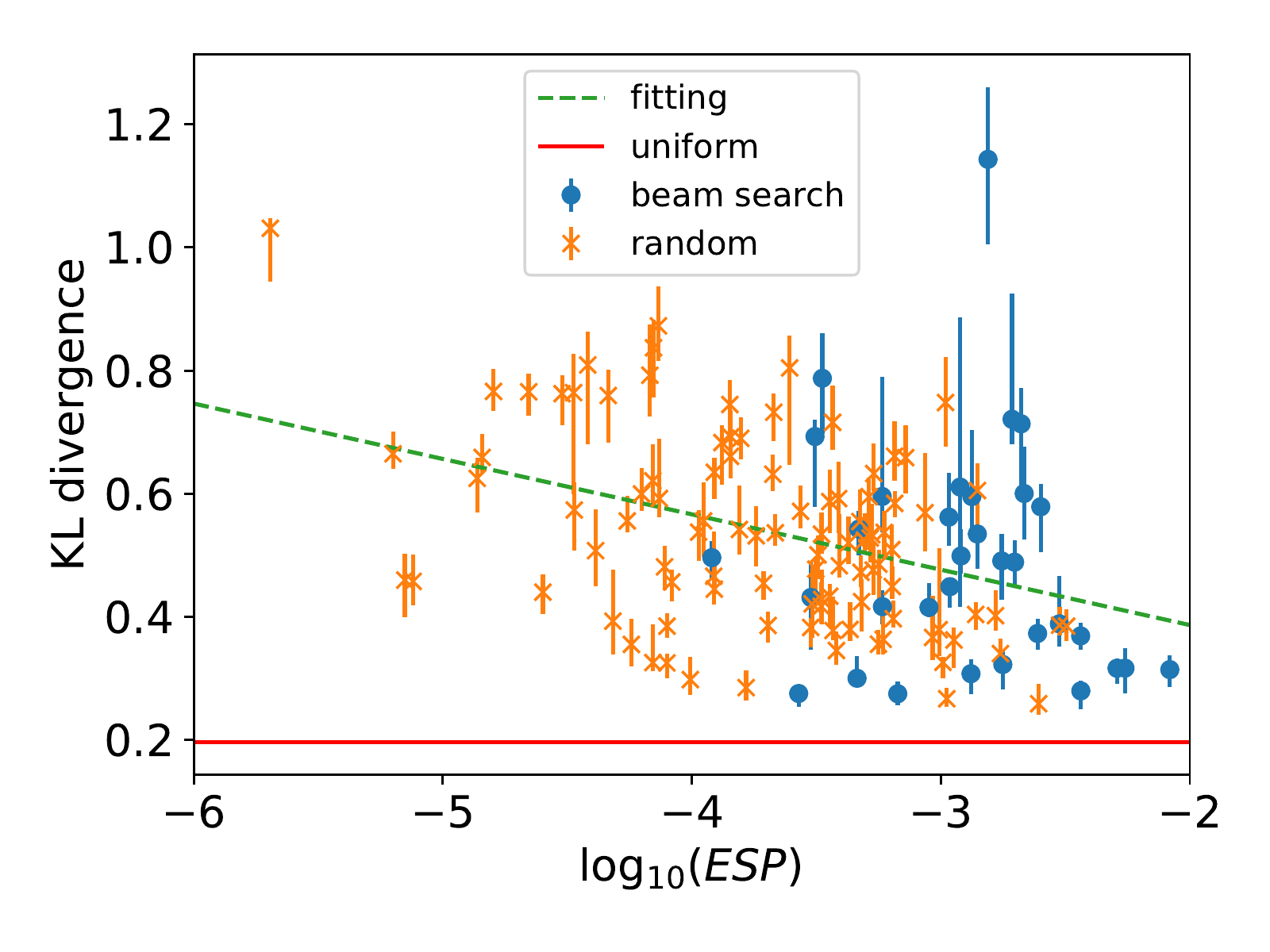}
        \subcaption{
            4-qubit adder
        }
        \label{fig:4-qubit-adder}
    \end{minipage}

    \caption{ 
    The relationship between ESP and KL divergence. In this experiment, we
    compiled a certain number of copies of the adder circuit with the beam search
    compiler and the random selection compiler (70 circuits for 1-qubit adder and 4-qubit
    adder, 100 circuits for 2-qubit). For each circuit, we executed a fixed
    number of experiments to observe fluctuation of KL divergence (5
    experiments for 1-qubit and 2-qubit adder, 10 experiments for 4-qubit). Each
    experiment ran the circuit 5000 times on IBM Q20 Tokyo and calculated KL
    divergence from the measurement results. The points in the plot show the
    median of the KL divergence for each circuit, and the error bars correspond
    to the maximum and the minimum KL divergence.
    }
\end{figure}

To test our compiler, we compiled Cucarro's ripple-carry adder
circuit~\cite{quant-ph/0410184}. Fig.~\ref{fig:adder}
shows the schematic. In the test bench, we first applied Hadamard gates for each
qubit to initialize the states, and then passed the qubits to the adder.
Finally, we measured the addition result and the carry-out in the computational
basis. We adopted the adder circuit because:
\begin{enumerate}
    \item the computation is simple to understand, and the ideal distribution is easy to compute; but
    \item the addition circuit is made of a complex combination of CNOT gates, and
    \item addition is a key component of many quantum algorithms, so its performance is inherently important.
\end{enumerate}

In this experiment, we compared the following two compilation algorithms:
\begin{enumerate}
    \item The beam search compiler.
    We used $B = 10000$ and $M = 1000$ as the parameters.
    \item A random selection compiler.
    Instead of ranking states via the score function, this compiler randomly picks a
    state $s$ from $S_i$ and set $S_{i+1}$ as a singleton $\{s\}$ in Algorithm
    \ref{alg:compile}. Also, we randomly selected one initial mapping as the
    starting point.
\end{enumerate}
Since the initial execution state is a superposition state where all values appear with
equal probability, the output of this circuit is identical to the distribution
of the sum of two rolls of dice.  Fig.~\ref{fig:pyramid} shows the ideal
distribution and observed distributions of two compilers.

We tested the adder circuits for input register sizes of one, two, and four qubits on IBM Q20 Tokyo
machine.  Table \ref{tab:compilation} shows the number of gates and the time required to compile.

Figures \ref{fig:1-qubit-adder}--\ref{fig:4-qubit-adder} show the
results.  The horizontal axis is the ESP of the compiled circuits.
Note that the horizontal axis is a logarithmic scale.  The vertical axis shows
the KL divergence between the observed distribution and the ideal distribution.
We sampled 5000 shots to compute one KL divergence for each experiment.
Moreover, to consider the variation of noises during the experiments, we ran
the same circuit several times.  Each point in the plot corresponds to the
median of KL divergence and the error bar shows the maximum and the minimum for
each compiled circuit. We have also included the KL divergence between the
ideal distribution and the uniform distribution as a guide.

First, we can see the points of beam search tend to concentrate in the right
half in all plots.  This shows beam search can indeed choose
realizations of gates such that ESP of the whole circuit improves dramatically.  Since the
horizontal axis is a logarithmic scale, our technique often results in one
magnitude higher ESP.

Next, Fig.~\ref{fig:1-qubit-adder} shows that most of the 1-qubit adder
experiments results in better output distributions than a uniform
distribution.  As the number of qubits increases and the circuit gets more
complicated, however, the KL divergence gets worse.  For a 2-qubit adder, only
half of the experiments performed better than a uniform distribution, and for a
4-qubit adder, no experiments did.  This tendency can be observed regardless of
the compilation algorithm.

Finally, we analyzed how KL divergence differs as the ESP of the compiled
circuit changes.  We draw the linear regression of KL and ESP as a dashed line
to each plot.  For the 1-qubit adder experiments shown in Figure
\ref{fig:1-qubit-adder}, there is a clear negative correlation between ESP and KL
divergence. Its correlation coefficient was $-0.475$.  Since lower KL
divergence means the output of the compiled circuit is closer to the ideal, the
negative trend implies that our approach of compiling a quantum circuit with
higher ESP actually improves the reliability of compiled circuits for the
1-qubit adder circuit.

When it comes to 2-qubit adder experiments (Fig.~\ref{fig:2-qubit-adder}),
the correlation coefficient of ESP and KL is $-0.0279$, which means there is
almost no correlation between ESP and KL.

Although there is also negative correlation between ESP and KL for the
4-qubit adder experiments, the absolute values of KL divergence are much worse
than that of even the uniform distribution.  So we can see the improvement of
KL divergence here, but the improvement is too small to make the computation
reliable.

%% file: adder.tex
\Qcircuit @C=0.9em @R=0.5em {
& \push{\ket{c}}  & \qw      & \mmaj & \qw   & \qw      & \qw   & \muma & \push{\ket{c}} \qw       & \qw \\
& \push{\ket{b0}} & \gate{H} & \gmaj & \qw   & \qw      & \qw   & \guma & \push{\ket{s_0}} \qw     & \meter \\
& \push{\ket{a0}} & \gate{H} & \gmaj & \mmaj & \qw      & \muma & \guma & \push{\ket{a_0}} \qw     & \qw \\
& \push{\ket{b1}} & \gate{H} & \qw   & \gmaj & \qw      & \guma & \qw   & \push{\ket{s_1}} \qw     & \meter \\
& \push{\ket{a0}} & \gate{H} & \qw   & \gmaj & \ctrl{1} & \guma & \qw   & \push{\ket{a_1}} \qw     & \qw \\
& \push{\ket{z}}  & \qw      & \qw   & \qw   & \targ    & \qw   & \qw   & \push{\ket{z + s_2}} \qw & \meter \\
}

%% file: conclusion.tex
\section{Future Work and Conclusion}
\label{sec:conclusion}

In this paper, we proposed two reliability metrics for quantum
gates and circuits: Estimated Success Probability for use during
compilation and KL divergence for assessing results.  Estimated
Success Probability is a composable measure of the quality of quantum
operations. The ESPs of qubit initialization, single qubit gates, CNOT
gates, and measurements are the error rates given by randomized
benchmarking.  We defined the ESP of a composite circuit as the
product of the ESPs of its components.  Due to its composability,
compilers can calculate ESP easily, so we adopted full-circuit ESP as
the optimization target of our compiler.

Since we cannot experimentally observe the ESP of circuits that end
with a superposition state just before measurement, including
important building blocks for other algorithms such as adders and the
Quantum Fourier Transform, we use KL divergence for comparing the
output distributions of experiments across compilation algorithms.

In lieu of full tomography, KL divergence is a useful intermediate
tool.

Experiments on path selection and circuit selection showed that using
ESP gives lower (better) KL divergence than random choice among
shortest-path candidates.  However, even in the relatively simple case
of selecting a two-hop path, our best success rate is only 70\%.

Despite the difficulty of path selection, our experiments showed that
our beam search-based compiler improves the ESP of a quantum
circuit. Our experiments with 1-qubit adders showed that the
improvement of ESP led to lower KL divergence, which means our
approach of optimizing ESP experimental can mitigate errors in NISQ
devices, at least for smaller circuits. However, as the circuit
becomes complex, the relationship between ESP and KL divergence
vanished, or KL got much worse than even the uniform
distribution. This behavior demonstrates the limit of current NISQ
computation, while highlighting the importance of compilation
aggressively focused on errors.

Although ESP is better than random selection, in some cases we can
make the optimal choice out of seven candidates 43\% of the
time. This is due to the fact that ESP and the physical reality are
divergent.

One shortcoming of our current approach is that it does not take into
account memory errors.  Due to the complexity of gate scheduling
within Qiskit, augmenting ESP via a gate-by-gate, qubit-by-qubit
insertion of Identity (``Wait'') gates complete with $T_1$ (energy
relaxation time) and $T_2$ (dephasing time) decoherence is a difficult
challenge.  We are considering methods to incorporate the execution
time and apply a blanket decoherence term.

We plan to test other circuits such as QFT for more
comprehensive benchmarking of the compiler. However, our evaluation
will not work for circuits which appear in the context of quantum
supremacy because our KL divergence based evaluation requires deriving
the ideal distribution with a classical computer
beforehand. Therefore, we need a method for estimating KL divergence
to better generalize our approach.  Additionally, we need a method for
extrapolating from KL divergence to the prospects of seeing the
expected quantum interference patterns that drive quantum algorithms;
this will involve assessing sign errors as well as bit flip errors.

Finally, now that we have shown the value of error-aware compilation,
we hope to influence the design of the Qiskit compiler (and compilers
for other languages and systems) or have our code incorporated
directly into the standard release.  As our ability to correctly
predict success improves, the tools can also be used as part of an
evaluation of architectural tradeoffs between different types of
system couplers and qubit layouts, influencing the design of future
generations of quantum computers.

%% file: acknowledgement.tex
\section*{Acknowledgement}
The results presented in this paper were obtained in part using an IBM Q quantum computing system as part of the IBM Q Network.
The views expressed are those of the authors and do not reflect the official policy or position of IBM or the IBM Q team.